\documentstyle[epsfig]{article}

\def\<{\langle}
\def\>{\rangle}

\begin{document}
\section*{Destruction of quantum
coherence and wave packet dynamics}
\begin{center}
Gernot Alber\\
Abteilung f\"ur Quantenphysik, Universit\"at Ulm, D-89069 Ulm,
Germany
\\
(to be published in {\em The Physics and Chemistry of Wave Packets},
edited by J. A. Yeazell and T. Uzer (Wiley, N. Y.))
\end{center}

The development of short, powerful laser pulses
and of sophisticated
trapping techniques within the last few years has
stimulated numerous theoretical and experimental
investigations on the dynamics of wave packets in
elementary, material quantum systems. 
These wave packets are non stationary, spatially
localized quantum states which are 
situated on the border between the microscopic and macroscopic
domain.  A detailed understanding of their dynamics is essential
for our conception of quantum mechanics and of its connection
with classical mechanics. So far the 
interplay between classical and quantum mechanical aspects
of their dynamics have been investigated in Rydberg systems
(Alber and Zoller 1991),
in molecules (Garraway and Suominen 1995, 
Sepulveda and Grossmann 1996),
in clusters (Knospe and Schmidt 1996) and in
nano structures (Koch et al. 1996). These studies
have concentrated  
mainly on semiclassical aspects which may be attributed to the
smallness of the relevant de Broglie wave lengths. Thereby quantum
aspects still manifest themselves in interferences
between probability amplitudes which are
associated with various
families of classical trajectories.
However, for a comprehensive understanding of the emergence of
classical behavior also a detailed
understanding of the destruction of quantum coherence
is required. Typically this destruction of coherence
arises from external stochastic forces or environmental influences
which cannot be suppressed.
Though by now many aspects of the coherent dynamics of these
wave packets
are understood to a satisfactory degree still scarcely anything
is known about the influence of destruction
of quantum coherence.

The main aim of this article is to discuss 
characteristic physical phenomena which govern the destruction
of quantum coherence of material wave packets.
For systematic investigations on this problem it is advantageous
to deal with physical systems in which
wave packets can be prepared and detected in a controlled
way and in which the mechanisms causing the destruction
of quantum coherence can be influenced to a large extent.
Rydberg atoms (Seaton 1983, Fano and Rau 1986)
are paradigms of elementary quantum systems
which meet these requirements. 
The high level density of Rydberg states close to an ionization
threshold is particularly convenient for the experimental
preparation of spatially localized electronic wave
packets by coherent superposition of energy eigenstates
(Alber and Zoller 1991).
Furthermore, the dynamics of electronic Rydberg wave packets
exhibits universal features which apply to atomic and molecular
Rydberg wave packets as well as to Rydberg wave packets in more
complex systems such as clusters. This dynamical universality might
be traced back to the fact that almost over
its whole classically
accessible range the dynamics of a Rydberg electron is governed
by the Coulomb potential of the positively charged ionic core.
This universality together with the fact that Rydberg systems
are amenable
to a systematic theoretical description with the help of 
semiclassical methods 
makes them attractive for theoretical investigations.
In recent years many detailed investigations
have been performed concerning
various fundamental aspects of the coherent
dynamics of Rydberg wave packets, such as
the influence of core scattering processes
(Alber 1989, Dando et al. 1995, H\"upper et al. 1995),
the connection between classical
bifurcation phenomena and quantum dynamics
(Beims and Alber 1993, 1996, Main et al. 1994),
the influence of the stimulated light force on the atomic center of
mass motion (Alber 1992, Alber and Strunz 1994) or
the influence of electron correlations on wave packet dynamics
in laser-induced two-electron excitation processes
(Hanson and Lambropoulos 1995, Zobay and Alber 1995,
van Druten and Muller 1995).

The dynamics of Rydberg electrons 
is governed by characteristic features 
which greatly influence the way in which they
can be affected
by external stochastic forces or environments. 
Most notably, Rydberg electrons
can be influenced by laser fields of moderate intensities
and by their statistical properties
only in a small region around the atomic nucleus
(Giusti and Zoller 1987). 
Furthermore, Rydberg systems are characterized by unique threshold
phenomena which result from the
infinitely many bound states and from the continuum states
converging towards an ionization threshold.
In addition, radiative decay rates of Rydberg states
are so small
that in typical situations of current
experimental interest the direct influence 
of radiative damping can be neglected.
However, the dissipative influence of radiative decay might
become significant, if Rydberg systems interact with intense laser
fields.
In order to demonstrate characteristic physical phenomena
governing the destruction of quantum coherence of electronic
Rydberg wave packets 
in the subsequent discussion 
two stochastic mechanisms will be
considered in detail,
namely radiative damping which is mediated by
electron correlations between a Rydberg wave packet and
a resonantly excited, tightly bound core electron
and fluctuations of laser fields.

The investigation of 
radiative damping mediated by electron correlation
effects is motivated by the recently revived interest
in laser-induced two electron excitation processes
(Jones and Bucksbaum 1991, Stapelfeldt et al. 1991,
Robicheaux 1993, Grobe and Eberly 1993,
Hanson and Lambropoulos 1995,
Zobay and Alber 1995, van Druten and Muller 1995).
Non-resonant laser-induced excitation processes in which
two valence electrons of an atom, e.g. an alkaline earth atom,
are excited simultaneously have already been playing
an important role in spectroscopy for a long time (Gallagher 1994).
Typically thereby one
of the valence electrons is excited into a Rydberg state
and the other one into a tightly bound core state.
Due to the availability of intense laser light sources recently
also cases have become accessible experimentally
in which both of these electrons
are excited resonantly so that the influence of the laser field
can no longer be treated with the help of perturbation theory.
The resulting strong modifications
of the electron-correlations may give rise to
interesting novel phenomena.
If the Rydberg electron is prepared in a wave packet state
these coherent
laser-modified electron correlations
may even lead to an almost complete suppression of
autoionization (Hanson and Lambropoulos 1995).
In the subsequent discussion it will be demonstrated that
these coherent effects are rather
sensitive to the destruction of coherence
which is caused by radiative
decay of the tightly bound, excited core electron.

Due to the inherent stochastic nature of laser light the investigation
of optical excitations of atoms or molecules by fluctuating laser
fields is one of the central problems of laser spectroscopy. 
So far research on this problem has concentrated predominantly
on laser-induced 
excitation of isolated energy eigenstates (Agarwal 1976,
Dixit et al 1980, Vemuri et al. 1991).
By now this special class of excitation processes
is understood to a satisfactory degree.
Despite these successes
so far scarcely anything is known about the effect of laser
fluctuations on 
optical excitation processes in which 
the level density of the resonantly excited states is large
and in which wave packets are prepared.
A paradigm in this respect is the laser-induced
excitation of Rydberg and continuum
states close to an ionization threshold
which typically leads to the preparation of an 
electronic Rydberg wave packet. 
This physical system is 
well suited for
investigating fundamental aspects
of the destruction of quantum coherence on wave packet dynamics.
In the subsequent discussion it will be demonstrated that this
fluctuation-induced destruction of quantum coherence together with
the peculiar threshold phenomena of Rydberg systems leads to a
variety of novel phenomena.
One of these generic effects is stochastic ionization
which manifests itself in a characteristic scenario of non-exponential
decays (Alber and Eggers 1997).

This paper is organized as follows:
In section 1 basic theoretical concepts
for describing the dynamics of Rydberg
electrons in laser fields are summarized.
This section focuses on coherent dynamical
aspects which
can be described conveniently with the help of semiclassical methods.
Within this framework quantum aspects manifest themselves in the
interference between probability amplitudes which are associated with
those classical trajectories along which probability is transported. 
In section 2 recent theoretical
work on the destruction of quantum coherence
in wave packet dynamics is reviewed.
Characteristic phenomena are exemplified by considering two 
dissipative mechanisms in detail.
In section 2.1 the influence of radiative damping on laser-induced
two-electron excitation processes is investigated. Effects of laser
fluctuations on the dynamics of electronic Rydberg wave packets are
discussed in Sec. 2.2.

\section{Coherent dynamics of Rydberg electrons
- general theoretical concepts}

In this section a brief review of general theoretical concepts
is presented which are useful for the description of the dynamics
of Rydberg electrons.
These concepts have already
been used successfully to
describe various aspects of the coherent dynamics of electronic
Rydberg wave packets.
Thereby we shall concentrate mainly on cases of recent
experimental and theoretical interest in which a weakly bound
Rydberg electron interacts with a laser field and additional
weak electric and/or magnetic fields
(Alber 1989, Dando et al. 1995, H\"upper et al. 1995,
Moser et al. 1997).
Throughout this review
Hartree atomic units will be used for which
$e=\hbar=m_e=1$ ($e$ and $m_e$ are the electronic charge and mass,
respectively).

Rydberg electrons are atomic or molecular electrons whose dynamics
is dominated by highly excited energy eigenstates
close to an ionization
threshold. In the simplest possible case the energies of these
Rydberg states are given
by the well known relation $\epsilon_n = -1/[2(n - \alpha)^2]$
(Seaton 1983, Fano and Rau 1986).
Thereby the quantum defect
$\alpha$ is approximately energy independent for energies sufficiently
close to the ionization threshold at energy $\epsilon = 0$.
In typical optical excitation processes
only Rydberg states with small values of the angular momentum $l$
are excited, i.e. $l\ll n$.
These Rydberg states of low angular momenta are essentially de-localized
over the whole space which is classically accessible to them, i.e.
$(l + 1/2)^2 < r < 1/\mid \epsilon_n \mid$. ($r$ denotes the radial
distance of the electron from the nucleus measured in units of the
Bohr radius $a_0 = 5.29 \times 10^{-11} {\rm m}$.)

If a Rydberg electron interacts with a laser field of moderate
intensity and with a weak, static electric and/or
magnetic field one can distinguish three characteristic spatial
regimes:
\\
\\
(1) {\em The core region:} ($0 <r < O(1)$)
\\

It extends a few Bohr radii around the atomic nucleus.
Inside this core region 
Rydberg electrons of low angular momenta which are able to
penetrate this core region interact with
all other atomic core electrons. These interactions
lead to characteristic electron correlations effects such as
autoionization and channel coupling.
Quantitatively these effects can be described by
quantum defect parameters 
which are approximately energy independent close to an ionization
threshold (Seaton 1983, Fano 1986, Aymar et al. 1996).

If a Rydberg electron of low angular momentum interacts with a 
laser field of moderate intensity, whose electric
field strength is given by
\begin{equation}
{\bf E}(t) = {\bf E}_0 e^{-i\omega t} + c.c.,
\end{equation}
two major effects take place.
Firstly, the Rydberg electron experiences an intensity dependent
ponderomotive energy shift of magnitude
$\delta \omega_p = \mid {\bf E}_0\mid^2/\omega^2$.
This energy shift is independent of the energy of the Rydberg electron
and may thus be interpreted as an energy shift of the ionization 
threshold.
Secondly, all other dominant energy exchange processes between 
a Rydberg electron and the laser field are localized 
within a region typically extending a few Bohr radii around the 
atomic nucleus. 
This localization of the electron-laser coupling inside the
core region relies on two sufficient conditions, 
namely moderate
laser intensities and sufficiently high
laser frequencies preferably in the optical frequency domain (Giusti
and Zoller 1987).
Thereby 
laser intensities are considered to be moderate provided
the stationary oscillation amplitude $\alpha_{osc}$
of an electron in the laser field (in the absence of the Coulomb
potential of the ionic core) is significantly less than the
extension of the core region, i.e.  
\begin{equation}
\alpha_{osc} = \mid {\bf E}_0\mid/\omega^2 \ll 1.
\label{alpha}
\end{equation}
Furthermore, in this context laser frequencies $\omega$
are considered to be high, if they are much larger than the
inverse classical Kepler period $T_n$ of the Rydberg electron, i.e.
$\omega T_n \gg 1$ with $T_n = 2\pi(n - \alpha)^3$.
Classically speaking at these high laser frequencies it is only
in a region close to the nucleus that the acceleration of a Rydberg
electron is sufficiently large that an appreciable energy exchange
of the order of $\Delta \epsilon \approx \omega$
can take place between the laser field and the Rydberg electron
(compare also with Eq.(\ref{Bohr})).
As a consequence the interaction of a
Rydberg electron with a laser field of
moderate intensity and sufficiently high
frequency is completely different from its
interaction with a microwave field whose frequency is comparable
with its classical Kepler frequency $1/T_n$.
Even if the field strength of such a microwave field is small
in the sense that $\alpha_{osc}\ll 1$, the small frequency of the
microwave field implies that energy can be exchanged with
the microwave field essentially at any
distance of the Rydberg electron from the atomic nucleus.
\\
\\
(2){\em The Coulomb region:}($O(1)< r < a$)
\\

Outside the core region the dynamics of a
highly excited Rydberg electron is dominated by the $1/r$
Coulomb potential of the positively charged ionic core.
If the Rydberg electron is influenced by a weak external
electric or magnetic field this is only valid for distances
of the Rydberg electron from the nucleus which are smaller
than the critical
distance $a \gg 1$ at which the external potentials
are no longer negligible. If this critical distance is located inside
the classically accessible region, i.e.
$a < 1/\mid \epsilon_n \mid$, these external fields influence the
dynamics of the Rydberg electron significantly.
\\
\\
(3){\em The asymptotic region:} ($1\ll a < r$)
\\

In the asymptotic region the influence of weak external fields 
is as important as the Coulomb force originating from the positively
charged ionic core. In general, in this region
the resulting dynamics of the Rydberg
electron is complicated by the fact that
its classical dynamics
is no longer integrable and exhibits signatures of chaos.
\\

In each of these characteristic
spatial regimes different approximations
can be applied for the dynamical description of the Rydberg electron.
All photon absorption and emission processes and all
electron correlation effects which take place inside the
core region have to be described quantum mechanically.
As the Bohr radius is small in comparison with
the extension of the Coulomb region and
of the asymptotic region, 
outside the core region
the dynamics of a Rydberg electron
can be described with the help of
semiclassical methods.

Starting from these elementary considerations a systematic
theoretical description of Rydberg electrons can be developed
which is based on a synthesis of semiclassical methods and of concepts
of quantum defect theory (Alber 1989, Alber and Zoller 1991).
Thereby solutions of the Schr\"odinger equation
which are valid inside the core region and at the boundary to the
Coulomb region have to be matched to semiclassical 
wave functions which are valid in the Coulomb region and in the
asymptotic region.
The values
of the wave function at the border between
the core region and the Coulomb region
are determined by the solution
of the Schr\"odinger equation inside the core region.
Within the framework of quantum defect theory these values 
are determined by approximately energy independent
quantum defect parameters.
These quantum defect parameters originate from two
different types of interactions, namely electron correlation effects
and laser-induced photon absorption and emission processes. 
For moderate laser intensities and sufficiently high frequencies
these
latter type of processes give rise to intensity dependent
quantum defects. Thus, in the simplest case of
a one-channel approximation, for
example,
these interactions inside the core region can be characterized
by a complex quantum defect of the form (Alber and Zoller 1988)
\begin{equation}
\mu = \alpha + i\beta.
\label{complex}
\end{equation}
The real part of this quantum defect defines the energies of the
Rydberg electron
in the absence of the laser field, i.e.
$\epsilon_n = -1/[2(n - \alpha)^2]$.
The imaginary part $\beta$ describes the influence of laser-induced
transitions of the Rydberg electron into continuum states well
above threshold. In lowest order of perturbation theory it is given by
\begin{equation} 
\beta = \pi\mid
\langle \epsilon=\omega
\mid {\bf d}\cdot {\bf E}_0 \mid \epsilon = 0\rangle
\mid^2
\label{imag}
\end{equation} 
with ${\bf d}$ denoting the atomic dipole operator.
For hydrogen and linearly polarized laser light, for example,
this imaginary part of the quantum defect can be
evaluated approximately with the help of the Bohr correspondence
principle.
According to this principle the dipole matrix element entering
Eq.(\ref{imag}) is approximated by a Fourier coefficient
of the classical trajectory of a Rydberg electron 
of energy $\epsilon = 0$
(Landau and Lifshitz 1975), i.e. 
\begin{equation}
\langle \epsilon=\omega
\mid {\bf d}\cdot {\bf E}_0 \mid \epsilon = 0\rangle
=  \frac{1}{2\pi}\int_{-\infty}^{\infty}
dt e^{i\omega t} {\bf x}(t)\cdot {\bf E}_0=
\frac{6^{2/3}}{2\pi\sqrt{3}}\Gamma(2/3) \omega^{-5/3}\mid {\bf E}_0\mid.
\label{Bohr}
\end{equation}
($\Gamma(x)=\int_0^{\infty}du u^{x-1}e^{-u}$
denotes the Euler gamma function.)
Thereby ${\bf x}(t)$ describes the parabolic classical trajectory 
of an electron which moves in the Coulomb field of the nucleus 
with energy $\epsilon = 0$. Consistent with the previous qualitative
discussion the $\omega^{-5/3}$-dependence in Eq.(\ref{Bohr})
demonstrates that the dominant contribution to this dipole matrix
element originates from a spatial region around the nucleus with 
a size of the order of $r_c \approx \omega^{-2/3}$. This characteristic
size $r_c$ is the distance a classical
electron of (asymptotic)
energy $\epsilon = 0 $ can depart from the nucleus during
the relevant photon absorption time $t_{photon} \approx 1/\omega$.

In the Coulomb and asymptotic region the quantum mechanical state
can be determined semi classically. In order to make these
ideas more precise let us consider the
general form of the 
semiclassical solution of the
time independent Schr\"odinger equation which is valid 
in the Coulomb and
asymptotic region. It has the general form (Maslov and Fedoriuk 1981,
Delos 1986)
\begin{eqnarray}
\psi(\epsilon, {\bf x}) &=&
\sum_j
\varphi(\epsilon,{\bf y}_j)
\sqrt{\frac{J(0,{\bf y}_j)}{\mid J(t_j,{\bf y}_j)\mid}}
e^{i[S_j(t_j,{\bf y}_j) - \mu_j(t_j)\pi/2]}.
\label{semi}
\end{eqnarray}
This wave function is determined by two different types
of quantities, namely the probability amplitude
$\varphi(\epsilon,{\bf y})$ of finding the electron at position
${\bf y}$ on the boundary between the
core region and the Coulomb region and by quantities which describe
the classical motion of the Rydberg electron outside the core region
(compare with Fig. 1).
The probability amplitude $\varphi(\epsilon, {\bf y})$ is determined
by the quantum defect parameters which describe the
electron correlations and the electron-laser interaction inside
the core region. According to Eq.(6)
the probability amplitude $\psi(\epsilon,{\bf x})$ of finding the
electron at position ${\bf x}$ outside the core region is also
determined by properties of all those classical trajectories $j$
which start at the boundary between the core
region and the Coulomb region
at position ${\bf y}$ 
and reach the final point ${\bf x}$ at any `time' $t$.
In this context the variable $t$ represents a curve parameter and not
a physical time. Together with the initial positions ${\bf y}$ the
curve parameter $t$
constitutes a global coordinate system for the family of classical
trajectories which leave the core region and which form a Lagrangian
manifold (Maslov and Fedoriuk 1981, Delos 1986).
The important classical properties of trajectory $j$
which determine $\psi(\epsilon,{\bf x})$ are:
\begin{enumerate}
\item
its classical action (eikonal) $S_j(t_j,{\bf y}_j)$,
\item
the determinant of its Jacobi field
$$J(t_j,{\bf y}_j) = \frac{dx_1\wedge dx_2\wedge dx_3}
{dt\wedge dy_1\wedge dy_2}\mid_j$$
 which characterizes
its stability properties, and
\item
its Maslov index $\mu_j(t_j)$ which characterizes the number of
conjugate
points and their multiplicity.
\end{enumerate}

According to this
general theoretical approach it is apparent that Rydberg systems
differ from one another only as far as their dynamics inside
the core region is concerned. This part of the dynamics
can be described generally
by a few quantum defect parameters. Thus Rydberg systems
exhibit universal behavior and 
the quantum defect parameters characterize the associated
universality classes.
Furthermore, the 
semiclassical analysis of the dynamics of the Rydberg electron
in the Coulomb region and in the asymptotic region implies that
probability amplitudes describing atomic transitions
between an initial and a final state
can be represented as a sum of
contributions which are associated with all possible classical paths
(including their multiple returns)
which connect the regions of support
of the initial and the final state.
In particular, if the
dominant contribution of a transition amplitude originates
from the core region, for example,
it is all classical paths
which start and end inside the core region which are relevant
for the theoretical description.
On the basis of this combination of methods of quantum defect theory
with semiclassical path representations for relevant quantum
mechanical transition amplitudes many aspects of the
coherent dynamics of electronic Rydberg wave packets have already been
described successfully (Beims and Alber 1993, 1996, Alber et al. 1994,
Zobay and Alber 1998).

\section{Dissipative dynamics of electronic Rydberg wave packets}

So far in the context of wave packet dynamics of material
particles the investigation of
dissipative and stochastic influences which destroy
quantum coherence has not received much attention. Definitely,
to some extent this may be attributed to the complications
arising from the high level densities 
which have to be taken into account for a proper theoretical
description of wave packet dynamics. In general they turn
the solution of master equations for the relevant density operator
into a difficult mathematical and numerical problem.
Electronic wave packets in
Rydberg systems are an extreme example of this kind
due to their almost macroscopic size and the infinitely high level
density of Rydberg states at an ionization threshold. 
In the subsequent discussion it will be demonstrated that a combination
of the semiclassical methods discussed in Sec. 1 together with stochastic
simulation methods constitutes a powerful theoretical approach for
describing many aspects of the destruction of quantum coherence
in wave packet dynamics.
In addition,
this theoretical approach
offers insight into the intricate interplay between the semiclassical
aspects of the dynamics of a Rydberg electron outside the core region
and its coupling to the radiation field inside the core region.
In the subsequent sections two types of physical processes will
be discussed in detail by which this coupling to the radiation field
can destroy the quantum coherence of an electronic wave packet, namely
spontaneous emission of photons and the intrinsic fluctuations of a
laser field.
Motivated by the recent interest
in laser-induced two-electron excitation
processes, in Sec. 2.1
characteristic effects of radiative damping are explored
which are mediated by the correlation between an electronic
Rydberg wave packet and
a resonantly excited, tightly bound core electron.
In Sec. 2.2 it is demonstrated that 
as a result of 
the peculiar
threshold properties of Rydberg systems 
the destruction of quantum
coherence which is brought about by a fluctuating laser field
gives rise to a variety of novel phenomena.

\subsection{Radiative damping mediated by electron correlations}

Due to the long radiative life times of Rydberg states
(radiative life times scale as $(n - \alpha)^3$ (Gallagher 1994))
the direct influence of spontaneously
emitted photons 
is negligible under typical laboratory situations.
However, destruction of quantum coherence
originating from radiative damping might become
significant in cases in which more than one atomic or
molecular electron is excited
resonantly by a laser field. In such cases the influence of
a photon which is emitted spontaneously by one of these excited
electrons can influence another excited Rydberg
electron via electron correlation effects. 
Isolated core
excitation (ICE) processes (Cooke et al. 1978)
are a particular class of laser-induced
two-electron excitation processes which has 
received considerable attention recently.
In the following it is demonstrated that
in these types of excitation processes the 
dissipative
influence of radiative damping mediated by electron correlations
may influence the dynamics of electronic wave packets
significantly.

ICE excitation processes have been studied extensively in the
alkaline earth elements as the corresponding singly-charged ions 
are excited easily with laser fields in the optical or near-uv regime.
In Fig. \ref{ICE}
a typical laser-induced ICE process is shown schematically
for a magnesium atom.
In a first step, the atom is excited from its $|3s^2\rangle$
ground state to a Rydberg state $|3snd\rangle$ by two-photon
excitation. After this excitation process the atom consists of
the Mg$^+$(3s) ionic core and the $nd$-Rydberg electron which
tends to be located at large distances from the core.
By applying a second laser pulse tuned to a resonance of the Mg$^+$
ion
the remaining core electron is excited, e.g. to the $3p$-state of the
ionic core.
The direct influence of the laser field
on the highly excited Rydberg electron is usually negligible
in comparison
to its interaction with the second, tightly bound valence
electron. But the laser field
influences the Rydberg electron indirectly by
electron correlation effects. Immediately 
after the core transition the Rydberg
electron experiences a "shakeup" by the
different short-range core potential to which it
has to accommodate. A quantitative measure for
the degree of this shakeup is given by
the difference between the quantum defects of
the two channels associated with the $3s$ and the $3p$-states of
the ionic core. 
The early work on ICE spectroscopy of alkaline earth elements
has concentrated on non-resonant core transitions which can
be described in lowest order of perturbation theory
with respect to the laser field
(Gallagher 1994).
Non-perturbative effects of laser fields
have become of interest 
only recently in connection with the development of powerful tunable
laser sources (Jones and Bucksbaum 1991, Stapelfeldt et al. 1991,
Robicheaux 1993, Grobe and Eberly 1993).
They are particularly
important in resonant core excitation processes in which one of the
laser fields induces Rabi oscillations of the ionic core. 
A variety of new coherent effects have been predicted 
theoretically in this context
(Robicheaux 1993, Hanson and Lambropoulos 1995,
Zobay and Alber 1995, van Druten and Muller 1995)
which rely on the coherent interplay between the Rabi oscillations
of the ionic core and the dynamics of an electronic Rydberg wave packet
which is influenced by these Rabi oscillations through the resulting
shakeup processes (For a review on these theoretical developments 
see Zobay and Alber 1998).
However, due to the possibility of spontaneous emission of photons
by the resonantly excited core electron
all these effects are expected to
be particularly sensitive to the resulting destruction of quantum
coherence.

In order to investigate these dissipative
effects in
detail let us consider a typical laser-induced two-electron
excitation process in an alkaline earth atom
as represented in Fig. \ref{8}.
It is assumed that the atom is prepared initially
in its ground state
$|g \rangle$. The atom is situated in a cw-laser field whose
electric field strength is given by
${\bf E}(t) = {\cal E}{\bf e} e^{-i\omega t} + c.c.$
and which 
is tuned near resonance with a transition of the positively charged
ionic core. Typically electron correlations imply that as long as
the atom remains in its initial state
$|g\rangle$ this laser field is well
detuned from any atomic transition. Thus the laser field
has negligible effect on
the atomic dynamics. But
as soon as an outer valence electron is excited
to Rydberg state close to an ionization
threshold the cw-laser field starts to induce
transitions between the two resonantly coupled states of the ionic core
which have energies $\epsilon_1$ and $\epsilon_2$, respectively.
Let us concentrate on a case in which one of the valence
electrons is excited coherently to Rydberg states by
a short and weak laser pulse with electric field
strength ${\bf E}_a(t) = {\cal E}_a (t) {\bf e}_a
e^{-i\omega_a t} + c.c.$ (Typically the pulse
envelope ${\cal E}_a(t)$ will be modeled by a Gaussian shape centered
around time $t_a$ with pulse duration $\tau_a$).
Thus a radial electronic Rydberg wave packet is prepared by this short
laser pulse (Alber and Zoller 1991).
This wave packet moves in the Coulomb field of the
positively charged ionic core. Whenever it penetrates the core region
it is shaken up by the Rabi oscillations of the resonantly driven core.
Furthermore,
whenever the core emits a photon spontaneously 
this emission process
will disrupt the relative phases of the electronic wave packet
and will thus destroy quantum coherence.
The dynamics of this electronic wave packet under the influence of
the Rabi oscillations of the ionic core can be investigated 
by typical pump-probe experiments, for example.

For the theoretical description of the resulting destruction
of quantum coherence 
one has to solve the corresponding optical Bloch equation
for the density operator of the two atomic valence electrons.
In the case depicted in Fig. \ref{8}, for example,
the optical Bloch equation is given by (Zobay and Alber 1996)
\begin{equation}
\dot{\rho}(t) = -i[H,\rho(t)] + \frac{1}{2}\{
[L,\rho(t) L^{\dagger}] + 
[L\rho(t), L^{\dagger}]\}.
\label{Bloch}
\end{equation}
Thereby the Hamiltonian
\begin{eqnarray}
H&=& \sum_{i,j=1,..,3} H_{i,j} + V_{ICE}
\end{eqnarray}
characterizes the coherent part of the dynamics.
The dynamics of the valence electrons is described
by the Hamiltonian 
\begin{equation}
H_{i,j} = ({\bf h}_{jj}\delta_{ij} +
{\bf V}_{ij} + \epsilon_{cj}\delta_{ij})|\Phi_i\rangle
\langle \Phi_j|
\label{H}
\end{equation}
with
\begin{equation}
{\bf h}_{jj} = -\frac{1}{2}\frac{d^2}{dr^2} +
\frac{l_j(l_j + 1)}{2r^2} - \frac{1}{r}.
\end{equation}
The short-range potential ${\bf V}_{ij}$ describes 
electron-correlation effects originating from the residual core
electrons (Aymar et al. 1996).
In ICE transitions
the angular momentum $l$
of the excited Rydberg electron is conserved
to a good degree of approximation, i.e.
$l_1=l_2=l$ (Gallagher 1994).
In the rotating wave approximation
the channel thresholds 
$\epsilon_{cj}$ are given by
$\epsilon_{c1} = \epsilon_1$,
$\epsilon_{c2} = \epsilon_2 - \omega$, 
$\epsilon_{c3} = \epsilon_3 - \omega$. 
The operator
\begin{equation}
V_{ICE} = -\frac{1}{2}\Omega(|\Phi_2\rangle \langle \Phi_1| +
|\Phi_1\rangle \langle \Phi_2|)\otimes {\bf 1}_r
\label{VICE}
\end{equation}
describes the laser-induced core transitions between the core states
$\mid \Phi_1\rangle$ and $\mid \Phi_2\rangle$ and $\Omega$
is the Rabi frequency originating from the cw-laser field.
The operator ${\bf 1}_r$ denotes the identity operator for the radial
coordinate of the Rydberg electron. Thus the role of the Rydberg
electron as a spectator becomes obvious from Eq.(\ref{VICE}).

The stochastic part of the dynamics of the density operator $\rho(t)$
is described by the Lindblad operator
\begin{equation}
L = \sqrt{\kappa}|\Phi_1\rangle \langle \Phi_2 | \otimes {\bf 1}_r
\label{Lind}
\end{equation}
which characterizes the radiative decay of the ionic core from
its excited
state to its ground
state by spontaneous
emission of photons with rate $\kappa$.

Due to the high level density of Rydberg states close to an
ionization threshold
and due to the presence of the
adjacent electron continuum usually severe problems arise, if
one tries to
solve the optical Bloch equation (\ref{Bloch}) numerically
by expanding the density operator $\rho(t)$ into a basis set of
atomic energy eigenfunctions. 
Many of these problems can be circumvented successfully by
combining the semiclassical methods as discussed in Sec. 1
with stochastic simulation methods (Zobay and Alber 1996).
Besides numerical advantages
this approach gives also direct insight into the classical
aspects of the dynamics of the Rydberg electron and the destruction
of quantum coherence caused by the radiative decay of the core.
Thereby the density operator is represented
by a (fictitious) ensemble of pure states which are
associated with definite numbers
of spontaneously emitted photons (Mollow 1975), i.e.
\begin{equation}
\rho(t) = \sum_{N=0}^{\infty} \rho^{(N)}(t),
\label{rhon}
\end{equation}
with the $N$-photon contributions
\begin{eqnarray}
\rho^{(N)}(t) &=& 
\int_0^{t} dt_N
\int_0^{t_N} dt_{N-1} \cdots
\int_0^{t_2} dt_1
|\psi(t|t_N,...,t_1)\rangle
\langle \psi(t|t_N,...,t_1)|.\nonumber
\end{eqnarray}
The time evolution of the $N$-photon states
$|\psi(t|t_N,...,t_1)\rangle$ 
is given
by
\begin{eqnarray}
|\psi(t|t_N,...,t_1)\rangle&=& 
e^{-iH_{{\rm eff}}(t - t_N)}\Theta(t - t_N)L
e^{-iH_{{\rm eff}}(t_N - t_{N-1})}\Theta(t_N - t_{N-1})L
\cdots\nonumber\\&&
Le^{-iH_{{\rm eff}}t_1}\Theta(t_1)|\psi(t=0)\rangle
\label{jump1}
\end{eqnarray}
with the effective (non-Hermitian) Hamiltonian
\begin{equation}
H_{{\rm eff}} = H - \frac{i}{2}L^{\dagger}L.
\label{Ham}
\end{equation}
($\Theta(x)$ is the unit step function.)
The physical interpretation of Eq.(\ref{jump1})
is straight forward. With each emission of a photon at one of the
$N$ random emission times
$t_1 \leq t_2 \leq ...\leq t_N$ the quantum state 'jumps' into
a new state by application of the Lindblad operator of Eq.(\ref{Lind}).
Between two successive jumps the state evolves according to the
Hamiltonian of Eq.(\ref{Ham}).
Thus the decomposition of Eq.(\ref{rhon})
may also be interpreted as an unraveling of the density operator
into contributions associated with all possible quantum jumps.
This decomposition of the density operator $\rho(t)$ offers significant
advantages in cases in which the number of spontaneously emitted photons
is small or in which the evaluation of the relevant pure states can
be simplified by the application of semiclassical methods. In particular,
it is possible to derive general
semiclassical path representations for the 
$N$-photon states of the
optical Bloch equation
(\ref{Bloch}). Thus all physical observables of interest can be expressed
as a sum of probability amplitudes which are associated with repeated
returns of a Rydberg electron to the ionic core. During its
motion under the influence of the Coulomb potential of the ionic core
photons may be emitted spontaneously by the laser-excited core
at any position of the Rydberg electron along its path. These photon
emission processes disrupt the coherent quantum mechanical
time evolution of the Rydberg electron.

As an example, let us consider 
a coherent process which has received considerable attention recently,
namely laser-induced stabilization against autoionization 
(Hanson and Lambropoulos 1995).
This effect is based on a synchronization between
the dynamics of the ionic
core, which performs Rabi oscillations,
and the dynamics of a laser-prepared
electronic wave packet. This effect may be understood as follows:
At the time of the preparation of the electronic Rydberg wave packet
by the short laser pulse the core is in its ground state.
If the mean Kepler period
$T_{orb}=2\pi(-2\overline{\epsilon})^{-3/2}$
($\overline{\epsilon}$ is the mean excited energy of the Rydberg
electron)
of this wave packet is chosen
equal to a multiple of the Rabi period $T_{Rabi}=2\pi/\Omega$
of the core, the Rydberg electron will encounter 
the core in the ground state at
each of its subsequent returns to the nucleus. 
As autoionization of a Rydberg electron can take place only inside
the core region (Seaton 1983, Fano and Rau 1986, Aymar et al 1996),
this implies that the effective autoionization rate of the electronic
wave packet will become much smaller than the autoionization rate
of the mean excited Rydberg state $\Gamma_{\overline{n}}$ in the absence
of the laser field.
In addition, it has been demonstrated
(Hanson and Lambropoulos 1995)
that this suppression of
autoionization is also accompanied by a reduction
of dispersion of the electronic wave packet.
This suppression of dispersion is
brought about by the Rabi-oscillating core
which acts like a quantum-mechanical shutter and effectively
cuts off the tails of the wave packet which arrive
at the nucleus out of phase with small probability.
As this stabilization against autoionization is based on the coherent
interplay between electron correlations and laser-induced Rabi
oscillations it is expected to be particularly sensitive against
the destruction of quantum coherence due to spontaneous emission
of photons by the ionic core.

In the presence of radiative decay of the ionic core the physical picture
is changed significantly. In the simplest case of synchronization, i.e.
 for $T_{orb}=T_{Rabi}$, the first photon will be emitted spontaneously
by the ionic core most probably at a time $(M+1/2)T_{Rabi}$ (with $M$
denoting any integer) because then the core is in its excited state
with high probability.
Due to the synchronization at these times
the electronic Rydberg wave packet
is close to the outer turning point of its Kepler orbit. The spontaneous
emission of a photon reduces the excited core to its ground state.
Therefore, at the subsequent return of the electronic
wave packet to the core at time $(M+1)T_{orb}$ the ionic core will
be in its excited state so that the Rydberg electron will autoionize
on a time scale of the order of $1/\Gamma_{\overline{n}}$.
Thus, the laser-induced stabilization against autoionization will be
destroyed. Typically, 
$\Gamma_{\overline{n}} \gg \kappa$
so that the Rydberg
electron will autoionize with high probability
long before the core can emit a second photon spontaneously.
Consequently, it is expected that the influence of the radiative
damping on this coherent stabilization
phenomenon can be described approximately
by taking into account only
the zero-and one-photon contributions of the
density operator $\rho(t)$.

The influence of radiative damping described above
manifests itself clearly in the
time-dependent autoionization rate $\gamma(t)$ into
channel three, for example,
which results
from the dynamics of the electronic Rydberg wave packet. An experimental
technique for measuring $\gamma(t)$
has been developed recently
(Lankhuijzen and Noordam 1996).
This time-dependent ionization rate $\gamma(t)$ can be
decomposed into $N$-photon
contributions with the help of semiclassical path representations, 
i.e. 
\begin{equation}
\gamma(t) = \sum_{N=0}^{\infty}
\int_0^{t} dt_N \cdots
\int_0^{t_2} dt_1
\gamma^{(N)}(t). 
\end{equation}
It is expected that the 
zero- and one-photon contributions (Zobay and Alber 1996)
\begin{eqnarray}
\gamma^{(0)}(t)&=&
 \frac{1}{2\pi}
(1 - e^{-4\pi{\rm Im}\mu_2})
\mid
\int_{-\infty + i0}^{\infty + i0} d\epsilon_1
e^{-i\epsilon_1 t} (0,1,0) {\bf O}
\sum_{M_1=0}^{\infty}
(e^{i2\pi\tilde{\nu}_1}\tilde{\bf \chi})^{M_1}\times
\nonumber\\
&&
e^{i2\pi\tilde{\nu}_1}
\tilde{\bf{\cal D}}_{g {\bf e}_a}^{(-)}
\tilde{\cal E}_a(\epsilon_1 - \epsilon)\mid^2,
\nonumber\\
\gamma^{(1)}(t)&=& 
(\frac{1}{2\pi})^3
(1 - e^{-4\pi{\rm Im}\mu_2})
\mid
\int_{-\infty + i0}^{\infty + i0} d\epsilon_1 d\epsilon_2
e^{-i\epsilon_2 (t - t_1)}e^{-i\epsilon_1 t_1}
(0,1,0) {\bf O}\times\nonumber\\
&&
\sum_{M_2=0}^{\infty}
(e^{i2\pi\tilde{\nu}_2}\tilde{\bf \chi})^{M_2}
{\tilde{\bf S}}^{(M_2,M_1)}_{2,1}
\sum_{M_1=0}^{\infty}
(\tilde{\bf \chi}e^{i2\pi\tilde{\nu}_1})^{M_1}
\tilde{\bf{\cal D}}_{g {\bf e}_a}^{(-)}
\tilde{\cal E}_a(\epsilon_1 - \epsilon)\mid^2
\label{gamma}
\end{eqnarray}
are dominant.
In Eqs.(\ref{gamma}) the laser-induced excitation by the
short laser pulse is characterized by the Fourier transform
of the pulse envelope
\begin{equation}
\tilde{\cal E}_a(\Delta \epsilon) =
\int_{-\infty}^{\infty} dt 
{\cal E}_a(t) e^{i\Delta \epsilon (t - t_a)}
\end{equation}
and by the $(3\times 1)$-column vector 
$\tilde{\bf{\cal D}}_{g {\bf e}_a}^{(-)}$ whose components
are the energy normalized photoionization dipole matrix elements
(Seaton 1983) into channels one, two and three. The dynamics of the
Rydberg electron under the influence of the Rabi oscillations of 
the ionic core are described by the $(3\times 3)$ scattering matrix
$\tilde{\bf \chi}$ and by the $(3\times 3)$ diagonal matrix
$e^{i2\pi\tilde{\nu}}$ with matrix elements
$(e^{i2\pi\tilde{\nu}})_{jj} = 
e^{2i\pi
[2(\tilde{\epsilon}_ {cj} - \epsilon)]^{-1/2}}
\Theta(\tilde{\epsilon}_ {cj} - \epsilon)$ ($j=1,2,3$).
All matrices and column vectors
with a tilde refer to the basis of photon-dressed core states
$|\tilde{\Phi}_j\rangle$ ($j=1,2,3$)
(Robicheaux 1993, Zobay and Alber 1995).
These dressed channel states
are related to the corresponding bare states
$|\Phi_j\rangle$
by the orthogonal transformation ${\bf O}$
which diagonalizes the laser-induced
core coupling, i.e.
\begin{eqnarray}
{\bf O}^{T} [ \epsilon_c - i\kappa/2|\Phi_2\rangle
\langle \Phi_2| -
\frac{1}{2}\Omega (|\Phi_2\rangle \langle \Phi_1| +
|\Phi_1\rangle \langle \Phi_2|)] {\bf O} = \tilde{\epsilon}_c.
\end{eqnarray}
Thereby the diagonal matrix $\tilde{\epsilon}_c$ ($\epsilon_c$)
contains the energies
of the dressed (bare) core states.
Thus the relations
$\tilde{\bf{\cal D}}_{g {\bf e}_a}^{(-)} = 
{\bf O}^T \bf{\cal D}_{g {\bf e}_a}^{(-)}$ 
and
$\tilde{\bf \chi} = {\bf O}^T {\bf \chi} {\bf O}$
hold with the bare photoionization
dipole matrix elements
$\bf{\cal D}_{g {\bf e}_a}^{(-)}$ 
and with the bare scattering matrix
\begin{equation}
\mbox{\boldmath $\chi$} =
\left( \begin{array}{ccc}
e^{2\pi i \mu_1} & 0 & 0 \\
0 & e^{2\pi i \mu_2} & \chi_{23} \\
0 & \chi_{32} & \chi_{33} 
\end{array} \right).
\label{scatt}
\end{equation}
The quantum defects of the bare
channels one and two are denoted $\mu_j$.
These channels have opposite parity and cannot be coupled by
electron correlation effects. 
The matrix elements $\chi_{23}$ and $\chi_{32}$ characterize the
configuration interaction between channels 2 and 3 which results
in autoionization of channel 2. The autoionization rate of
a Rydberg state of channel 2 with principal quantum number $n$ 
is related to the imaginary part of the quantum defect $\mu_2$
by $\Gamma_n = 2{\rm Im}(\mu_2)/[n - {\rm Re}(\mu_2)]^3$.

Eqs.(\ref{gamma}) are examples of semiclassical path representations
for the zero- and one-photon ionization rates
$\gamma^{(0)}(t)$ and $\gamma^{(1)}(t)$. Their physical interpretation
is straight forward:
After the initial excitation by the short laser pulse those fractions
of the electronic Rydberg wave packet which are excited into
closed photon-dressed core
channels return to the core region periodically.
The integers $M_1$ and $M_2$  count the numbers of these returns.
Between two successive returns the Rydberg electron acquires
a phase of magnitude 
$(2\pi\tilde{\nu})_{jj}$
while moving in the photon-dressed core channel $j$.
This phase equals the classical action of motion along a purely
radial Kepler orbit with zero angular momentum and energy
$\epsilon - \tilde{\epsilon}_{cj} < 0$.
Entering the core region the Rydberg electron is scattered
into other photon-dressed core channels by laser-modified
electron correlation effects which are described by the scattering
matrix $\tilde{\bf \chi}$. The ionic core can emit a photon
spontaneously at any time during the motion of the Rydberg electron.
Quantitatively this photon emission process is described the
the quantity
\begin{equation}
{\tilde{\bf S}}^{(M_2,M_1)}_{2,1} =
\int_{0}^{T_{M_1, M_2}} d\tau
e^{2i\pi\tilde{\nu}_2(1 - \tau/T_{M_1,M_2})}
(e^{-i\pi/2} \tilde{\bf L})
e^{2i\pi\tilde{\nu}_1\tau/T_{M_1,M_2}}
\label{S}
\end{equation}
in Eqs.(\ref{gamma}) with $T_{M_1,M_2} = t/(M_1 + M_2 + 1)$.
According to Eq.(\ref{S}) this spontaneous photon emission
by the ionic core can take place at any time $\tau$ between
two successive returns of the Rydberg electron to the core region.
At time $\tau$ the Rydberg electron has acquired a phase of magnitude
$(2\pi\tilde{\nu})_{jj}\tau/T_{M_1,M_2}$ in channel $j$.
The disruption of the phase of the Rydberg electron by this spontaneous
emission process is described by the action of the Lindblad operator
$\tilde{\bf L} = {\bf O}^T {\bf L} {\bf O}$. It also
leads to a
phase change of magnitude $(\pi/2)$.
After the completion of the photon emission process the Rydberg 
electron acquires an additional phase of magnitude 
$(2\pi\tilde{\nu})_{jj}(1 - \tau/T_{M_1,M_2})$ in the photon-dressed
core channel $j$ until it reaches the core region again.

A representative
time evolution of the autoionization rate $\gamma(t)$
is shown in Fig. \ref{10}.
The full curve in 
Fig. \ref{10}a has been obtained by numerical solution of the
optical Bloch equation (\ref{Bloch}) with the help of a conventional
basis expansion in atomic energy eigenstates.
The corresponding zero- and one-photon
contributions are also presented in Figs. \ref{10}b and \ref{10}c.
In Fig. \ref{10} the sum of zero- and one-photon contributions are
not plotted as they are indistinguishable from the numerical result
(full curve in Fig.\ref{10}a).
The chosen parameters
represent typical values realizable in alkaline earth experiments.
The comparison of $\gamma(t)$ (full curve of Fig.\ref{10}a)
 with the corresponding result in
the absence of radiative damping (dotted curve in Fig.\ref{10}a)
demonstrates that the influence of radiative damping is already 
significant at interaction times of the order of $T_{orb}$.
With the help of the zero- and one-photon contributions of
Eq.(\ref{gamma}) the dissipative influence of radiative damping
can be analyzed in detail.
As apparent from 
Fig. \ref{10}b, the zero-photon rate vanishes at integer multiples
of the mean Kepler period $T_{orb}$ because at these times the core
is in its ground state. The maxima of
Fig. \ref{10}b at times $(M + 1/2)T_{orb}$
originate from fractions of the electronic wave packet
which are close to the core at times when the core is in its
excited state. Also visible are typical revival effects
at times of the order of $25T_{orb}$.
The one-photon rate of
Fig. \ref{10}c exhibits maxima and minima at times
$MT_{orb}$ and $(M + 1/2)T_{orb}$.
These maxima indicate that the photon is emitted by the ionic core most
probably whenever the Rydberg electron is close to the outer
turning point of its classical Kepler orbit.
Thus, the core will be in its excited state when the Rydberg electron
returns to the nucleus so that autoionization will take place
with a high probability.

\subsection{Electronic wave packets in fluctuating laser fields}

The main aim of this section is to discuss characteristic
effects which govern the dynamics of a Rydberg electron in an 
intense and fluctuating laser field. It is demonstrated that for
moderate laser intensities (compare with Sec. 1 Eq.(\ref{alpha}))
a variety of novel, non-perturbative effects appear which influence
the long time behavior of Rydberg electrons significantly.
A generic consequence of the interplay between the peculiar
threshold phenomena of Rydberg systems and the destruction
of quantum coherence due to laser fluctuations is stochastic
ionization (Alber and Eggers 1997). It is demonstrated that
this process also implies an upper time limit on the applicability
of two-level approximations even in cases in which all characteristic
frequencies, i.e. Rabi frequencies and laser bandwidths, are small
in comparison with the Kepler frequency of a resonantly excited
Rydberg electron.

Nowadays laser fluctuations can be controlled to such a degree
that it is possible to realize various theoretical models
of laser radiation in the laboratory (Vemuri et al. 1991).
One of the most elementary theoretical models of laser radiation is
the phase diffusion model (PDM) (Haken 1970).
It describes approximately
the electric field produced by an ideal single
mode laser which is operated well above the laser threshold.
Thereby the electric field of a laser
is represented by a classical, stochastic
process
with well stabilized amplitude and a fluctuating phase, i.e.
\begin{equation}
{\bf E}(t) = {\bf E}_0 e^{i\Phi(t)}e^{-i\omega t} + c.c.~.
\end{equation}
The fluctuations of the phase $\Phi (t)$
are modeled by a real-valued Wiener process (Kl\"oden and Platen 1992),
i.e.
\begin{eqnarray}
&&M d\Phi (t) = 0,  [d\Phi(t)]^2 = 2b dt
\label{PDM}
\end{eqnarray}
Thereby
M indicates the mean over the statistical ensemble.
The PDM implies a Lorentzian spectrum of the laser radiation
with bandwidth $b$.

In order to investigate the influence of laser fluctuations on
the optical
excitation of Rydberg states close to an ionization threshold let
us consider the simplest possible case, namely
one-photon excitation from a tightly
bound initial state $|g\rangle$ with energy $\epsilon_g$.
In the dipole and rotating wave approximation
the Hamiltonian which describes this excitation process
is given by
\begin{eqnarray}
H(\Phi(t)) &=&
\epsilon_g |g\rangle \langle g| +
\sum_n \epsilon_n |n \rangle \langle n|  -\nonumber\\
&&
\sum_n (|n \rangle \langle g| \langle n|{\bf d}|g\rangle \cdot
{\bf E}_0 e^{i\Phi(t)}e^{-i\omega t} + {\rm h.c.}). 
\label{Hamilton}
\end{eqnarray}
In Eq.(\ref{Hamilton})
the index $n$ refers to Rydberg and continuum states.
The energies of the
excited Rydberg states are denoted $\epsilon_n$ and ${\bf d}$ is the
atomic dipole operator.
Let us assume for the sake of simplicity that
the excited Rydberg and continuum states can be described
with the help of quantum defect theory
in a one channel approximation (Seaton 1983). Thus they
are characterized by an approximately energy independent quantum defect
$\mu=\alpha + i \beta$. 
As has been explained in Sec. 1 (Eq.(\ref{imag}))
the imaginary part $\beta$ describes photon absorption from
the highly excited Rydberg states to continuum states well above
threshold.

For the description of non-perturbative aspects of this laser excitation
process one has to solve the time dependent
Schr\"odinger equation with
the stochastic
Hamiltonian (\ref{Hamilton}) (interpreted as a stochastic
differential equation of the
Ito type (Kl\"oden and Platen 1992))
together with the stochastic differential
equation for the phase (\ref{PDM}). 
It is the simultaneous presence of the Coulomb threshold
with its infinitely many
bound states and the continuum on the one hand and the laser fluctuations
on the other hand which makes this solution a highly nontrivial task.
Nevertheless, for the case of the PDM the resulting mathematical and
numerical problems
can be circumvented successfully (Alber and Eggers 1997).
Thus even analytical results can be derived
in the limit of long interaction times which
is  dominated by stochastic ionization of the Rydberg electron.
Thus, let us start from the equation of motion
for the mean values
$\rho_{n n'}(t)=
M\langle n\mid \psi(t)\rangle \langle \psi(t)\mid n'\rangle$,
$\rho_{n g}(t)=[\rho_{g n}(t)]^*=
Me^{-i\Phi(t)}
\langle n\mid \psi(t)\rangle \langle \psi(t)\mid g\rangle$ and
$\rho_{g g}(t)=
M\mid \langle g\mid \psi(t)\rangle \mid^2$
which can be combined to form a density operator $\rho(t)$
(Agarwal 1976).
From Eqs. (\ref{PDM}) and (\ref{Hamilton}) it can be shown 
that this density operator fulfills the master equation
\begin{equation}
\frac{d}{dt}\rho(t) = -i[H_{mod}, \rho(t)] + \frac{1}{2}
\{[L,\rho(t) L^{\dagger}] +
[L\rho(t), L^{\dagger}]\}.
\label{master1}
\end{equation}
Thereby the modified Hamiltonian $H_{mod}\equiv H(\Phi(t)\equiv 0)$
describes laser induced excitation of Rydberg states close to threshold
in the absence of phase fluctuations.
The destruction of quantum coherence which is brought
about by the laser fluctuations is 
characterized by the Lindblad operator
\begin{equation}
L = \sqrt{2b}|g\rangle \langle g|.
\label{Lind1}
\end{equation}
On the basis of this master equation Fourier representations can be
developed for the density matrix
elements whose kernels can be determined explicitly
with the help of quantum defect theory.
Thus all complications arising from the Coulomb threshold
are taken into account properly. 
These Fourier representations are useful
for numerical 
calculations of averaged transition probabilities which are highly
accurate even in the limit of long interaction times.
Furthermore, these representations are convenient starting
points for the derivation of analytical results.
Thus, the averaged initial state probability $\rho_{gg}(t)$,
for example, is given by (Alber and Eggers 1997)
\begin{eqnarray}
\rho_{gg}(t) &=&
\sum_{N=0}^{\infty}
\frac{1}{2\pi} \int_{-\infty + i0}^{\infty + i0} dz e^{-izt}
A_{gg}(z)[2bA_{gg}(z)]^{N} = \nonumber\\
&&
\frac{1}{2\pi} \int_{-\infty + i0}^{\infty + i0} dz e^{-izt}
A_{gg}(z)[1 - 2bA_{gg}(z)]^{-1}
\label{Agg}
\end{eqnarray}
with 
\begin{eqnarray}
A_{gg}(z) &=& U(z) + U^*(-z),
\label{kernel}\\
U(z)&=&\{
-C_1(z) + C_2(z) + \nonumber\\
&&i \sum_{{\rm Re}\tilde{\epsilon}_n < 0}
[1 - \frac{d}{dz}\Sigma^*(z_1 - z)]^{-1}
[z_1 - \overline{\epsilon} + ib - \Sigma (z_1)]^{-1}
\mid_{z_1=z+\tilde{\epsilon}^*_n}
\}\Theta(z)\nonumber
\end{eqnarray}
and with
\begin{eqnarray}
C_1(z) &=&
\frac{1}{2\pi(z + 2ib)}{\rm ln}
\frac{z - \overline{\epsilon} + i(b + \gamma/2)}
{-\overline{\epsilon} + i(\gamma/2 - b)},
\nonumber\\
C_2(z) &=&
\frac{1}{2\pi[z + i(\gamma + 2b)]}{\rm ln}
\frac{z - \overline{\epsilon} + i(b + \gamma/2)}
{-\overline{\epsilon} + i(\gamma/2 + b)}.
\end{eqnarray}
In the spirit of the discussion of Sec. 2.1. (compare
with Eq.(\ref{rhon}))
$\rho_{gg}(t)$ is represented as a sum of contributions
of all possible quantum jumps $N$ which can be induced
by the Lindblad operator of Eq.(\ref{Lind1}).
According to Eq.(\ref{Agg}) these contributions 
give rise to a geometric series which can be summed easily.
The sum appearing in Eq.(\ref{kernel})
extends over all dressed states $\tilde{\epsilon}_n$ of the
effective Hamiltonian $H_{\rm eff} = H_{mod} - i L^{\dagger}L/2$.
The mean excited energy is given by
$\overline{\epsilon} = \epsilon_g + \omega + \delta \omega $ with
$\delta\omega$ denoting the relative quadratic Stark shift
between the initial state $\mid g \rangle$ and the ponderomotive
shift of the excited Rydberg states (compare with the general
discussion in Sec. 1).
Besides the threshold contributions $C_1(z)$ and $C_2(z)$ 
the characteristic kernel $A_{gg}(z)$
is determined by the (resonant part of the
) self energy of the initial state $|g\rangle$, i.e.
\begin{eqnarray}
\Sigma(z)&=& \sum_n \frac{\mid \langle n | {\bf d}\cdot
{\bf E}_0|g\rangle \mid^2}{z - \epsilon_n} =
-i\frac{\gamma}{2} - i\gamma
\sum_{M=1}^{\infty}
(e^{i2\pi(-2z)^{-1/2}} \chi)^{M}.
\label{self}
\end{eqnarray}
This self energy is characterized by the laser-induced depletion rate
\begin{eqnarray}
\gamma&=&2\pi \mid \langle \epsilon = 0 |   
{\bf d}\cdot {\bf E}_0
|g\rangle \mid ^2
\end{eqnarray}
of the initial state $|g\rangle$ and by the scattering matrix element
\begin{eqnarray}
\chi&=& e^{i2\pi\mu}
\end{eqnarray}
which describes all effects arising from scattering
of the Rydberg electron by the ionic core and from photon absorption
(compare with Eq.(\ref{complex})).
The sum over $M$ in Eq.(\ref{self})
originates from
the multiple returns of 
the Rydberg electron to the core region
where the dominant contribution to the self energy comes from.
With each of these returns the Rydberg electron of energy $z < 0$
accumulates a phase of magnitude
$2\pi(-2z)^{-1/2}$
and with each traversal of the core region
it accumulates a (complex) phase of magnitude
$2\pi\mu$ due to scattering by the core
and due to photon absorption.
The laser-induced depletion rate $\gamma$, the imaginary part of
the quantum defect $\beta$ and the second order Stark shift
$\delta\omega$ 
describe the influence of the laser field
on the Rydberg electron.
As these quantities depend on the laser intensity
they are not affected by 
phase fluctuations of the laser field.

Master equations of the form of Eq.(\ref{master1})
with a self adjoint Lindblad operator are of general interest
as phenomenological models of continuous quantum measurement
processes (Braginsky and Khalili 1992).
In this context Eq.(\ref{master1}) would model 
excitation of Rydberg and continuum states close to an ionization
threshold by a classical, deterministic laser field
in the presence of continuous measurement of the initial state
$\mid g\rangle$. 
Thereby the inverse bandwidth $1/b$  would determine the mean time
between successive measurements.

Some qualitative aspects of the time evolution of an excited Rydberg
electron under the influence of a fluctuating laser field
are apparent from the contour plots of Figs. \ref{two}
and \ref{wave}
which refer to one-photon
excitation of a hydrogen atom by linearly polarized laser light with
$|g\rangle = |2s\rangle$. It is assumed that Rydberg states around
$\overline{n}=(-2\overline{\epsilon})^{-1/2} = 80$
are excited. According to the general discussion
in Sec. 1 (compare with Eq.(\ref{imag}))
the laser-induced transitions from the excited Rydberg states
to continuum states well above threshold are described
by an imaginary quantum defect with
$\beta = 0.00375\gamma$.

In Fig. \ref{two}a 
both the bandwidth of the laser field $b$ and the field-induced
depletion rate $\gamma$ of state $\mid g\rangle$ are assumed
to be small in comparison with the mean level spacing of the excited
Rydberg states, i.e. $b,\gamma \ll \overline{n}^{-3}$. Thus, one may
be tempted to
think that this excitation process can be described
well within the framework of a two-level approximation in which only
states $|2s \rangle$ and $|80 p\rangle$ are taken into account.
However,
Fig. \ref{two}a
 demonstrates that this expectation is only valid for sufficiently
small interaction times.
Indeed, the early stages of the excitation process are dominated
by Rabi oscillations of the electron between the initial and
the resonantly excited state.
These Rabi oscillations
are damped by the fluctuating laser field. An equilibrium
is attained for interaction times
$t\geq 1/b$ for which all coherence between the two resonantly
coupled states is
negligibly small and for which $\rho_{gg}(t) \approx
\rho_{\overline{n} \overline{n}}(t) \approx 1/2$.
This characteristic, well known two-level behavior is exemplified in
Fig. \ref{two}a
by the stationary probability distribution of the excited Rydberg state.
(The probability distribution of state $|g\rangle$
 which is localized in a region of a few Bohr radii around the
nucleus is not visible on the radial scale of Fig. \ref{two}a).
Fig. \ref{two}a  indicates that
for interaction times which are larger than a critical time $t_1$ 
this simple picture of the two-level approximation breaks down
and the probability distribution of the excited
Rydberg electron starts to spread towards larger distances
from the core. 
(here $t_1 \approx 5\times 10^{5}T$ with $T=2\pi\overline{n}^3$
denoting the mean classical orbit time).
Simultaneously the probability distribution becomes
more and more spatially de-localized with all nodes disappearing.
In order to obtain a more detailed understanding of this
diffusion-like process the time evolutions of
the initial state probability and of the ionization
probability are shown in Fig.\ref{two}b .
From Fig.\ref{two}b it is apparent that this 
spatial spreading of the Rydberg electron
is connected with a diffusion 
in energy space towards the ionization threshold.
At interaction times $t\geq t_c\approx 7\times 10^{9}T$ 
the Rydberg electron has reached the ionization threshold
and the ionization probability $P_{ion}(t)$ rises significantly
from a negligibly small value to a value close to unity.
Simultaneously
the initial state probability
$P_{gg}(t)$ starts to decrease faster.
This stochastic diffusion of the Rydberg electron which eventually
leads to ionization is a characteristic phenomenon brought about by
the fluctuations of the exciting laser field.
With the help of the theoretical approach presented above
this characteristic stochastic ionization process can be analyzed in
detail.
Thus it can be shown (Alber and Eggers 1997)
that the diffusion of the Rydberg electron
towards the ionization threshold
starts at time
\begin{equation}
t_1 = \frac{8}{\pi b\gamma T}
\label{t1}
\end{equation}
and eventually leads to stochastic ionization at interaction times
$t\geq t_c$ with
\begin{equation}
t_c = \frac{4\pi}{\sqrt{27}\gamma b}
[\frac{(\overline{\epsilon}^2 + 3(b^2 + \gamma^2/4)/4)^{3/2}}
{
\overline{\epsilon}^2 + b^2 + \gamma^2/4
}]^{1/2}.
\label{tc}
\end{equation}
The time evolution of $P_{gg}(t)$ is approximately given by
\begin{equation}
P_{gg}(t) = \frac{2}{\sqrt{\pi}}[2b\gamma T]^{-1/2} t^{-1/2}
\label{1}
\end{equation}
for $t_1 < t < t_c$ 
and crosses over to the power law
\begin{equation}
P_{gg}(t) = \frac{(\gamma + 2b)^2}{(2b\gamma\varphi/\pi)^2}
[\frac{\gamma b \Gamma^3(5/3)}
{27\pi(\overline{\epsilon}^2 + b^2 + \gamma^2/4)}]^{1/3}t^{-5/3}
\label{2}
\end{equation}
for interaction times $t > t_c$.
The variable $\varphi$ characterizes the distance of the mean
excited energy $\overline{\epsilon}$ from the ionization
threshold and is determined by
the relation
$-\overline{\epsilon} + i(b + \gamma/2) = Re^{i\varphi}$
($0\leq \varphi <\pi$). At times $t\geq t_c$
the ionization probability rises according to the
power law
\begin{equation}
P_{ion}(t) = 1 - \frac{\pi\Gamma(2/3)(\gamma + 2b)}{6b\gamma \varphi}
[\frac{\gamma b}
{\pi(\overline{\epsilon}^2 + b^2 + \gamma ^2/4)}]^{1/3} t^{-2/3}.
\label{ion}
\end{equation}
These approximate time evolutions are indicated by the dashed curves in
Fig. \ref{two}b.
The analytical results of Eqs.(\ref{t1}) and (\ref{tc}) explicitly
show how the critical times $t_1$ and $t_c$
for the breakdown of the two-level approximation and for stochastic
ionization
depend on the characteristic parameters
of the problem, namely the mean excited energy $\overline{\epsilon}$,
the laser bandwidth $b$ and the laser-induced depletion
rate of the initial state $\gamma$.

In  Fig. \ref{wave} 
both the laser bandwidth and the laser-induced
depletion rate of the initial state $|g\rangle$
are larger than the mean level spacing
$\overline{n}^{-3}$ of the excited Rydberg states. As in this case
the initial state is depleted by
the laser field in a time which is small in comparison with the
mean
Kepler period of the excited Rydberg states, i.e.
$1/\gamma \ll T=2\pi\overline{n}^{3}$,
an electronic Rydberg wave packet is prepared by power broadening
(Alber and Zoller 1988).
The initial stage of the preparation 
of this electronic wave packet by power broadening
manifests itself in an
approximately exponential decay of $P_{gg}(t)$ with
rate $\gamma$. The repeated returns of fractions of this wave packet
to the core region give rise to recombination maxima of $P_{gg}(t)$ which
occur roughly at multiples of the mean Kepler period $T$.
In the absence of laser fluctuations the non-perturbative time
evolution of such an  electronic wave packet under the influence of
a laser field is already well understood. In the completely
coherent case with each return
to the core region a fraction of the electronic wave packet can be 
scattered resonantly 
in the presence of the laser field
by  stimulated emission and
reabsorption of a laser photon accompanied by an electronic
transition to the initial
state $|g\rangle$ and back again.
This emission and reabsorption process of a laser photon
causes a time delay of the electronic
wave packet of the order of $1/\gamma$ with respect to
un-scattered fractions of the electronic wave packet. 
These repeated scattering processes lead to a
splitting of the original wave packet into many 
partially overlapping fractions.
In the completely coherent case
the interference of these overlapping fractions inside the core region
eventually give rise to a complicated time dependence of $P_{gg}(t)$
(Alber and Zoller 1991).

Characteristic
qualitative aspects of the time evolution of an electronic
wave packet in the presence of laser fluctuations are apparent from
Fig. \ref{wave}a.
Clearly, the initial stages of the time evolution are dominated
by the preparation of the electronic wave packet
and by its repeated returns
to the core region. However, at sufficiently long interaction times
eventually the spatially localized electronic
wave packet starts to spread out uniformly
over the whole classically accessible region.
Furthermore, this classical region starts to grow monotonically
with increasing interaction time.
Characteristic quantitative details of this time evolution are apparent
from Fig. \ref{wave}b. For sufficiently small interaction times
the familiar recombination maxima
of the repeated returns of the electronic
wave packet to the core region are clearly visible.
However,
as the coherence time of the laser field is small in comparison with
the mean Kepler period, i.e. $1/b \ll T$, interferences between
probability amplitudes which are associated with repeated returns to
the core region
are destroyed. Thus the details of the early stages of the time
evolution of this electronic wave packet appear to be much simpler than
in the completely coherent case. As a consequence of the diffusion
of the electronic wave packet at longer interaction times
the recombination maxima of $P_{gg}(t)$
disappear and merge into the power law of Eq.(\ref{1}).
At interaction times larger than $t_c$ stochastic ionization of the
Rydberg electron becomes significant and the power law decay of
$P_{gg}(t)$ crosses over to the decay law of Eq.(\ref{2}). Simultaneously
the ionization probability rises to a value close to unity
according to the approximate power law of Eq.(\ref{ion}).

In general stochastic ionization originating from laser fluctuations
will compete with other coherent ionization mechanisms such as
autoionization. As a consequence a
number of new interesting phenomena are expected to arise which are
not yet explored. In order to obtain first insights into basic
aspects of this competition let us generalize our previous model
to one-photon excitation of an autoionizing Rydberg series
(Eggers and Alber 1998). Thus,
it will be assumed that the laser excited autoionizing Rydberg
series can be described within the framework of quantum defect theory
in a two-channel approximation. In particular,
let us concentrate on a case in which the fluctuating laser field
excites Rydberg states 
close to an ionization threshold
of an excited state of the ionic core (channel one) which can autoionize
into channel two. For simplicity let us assume that direct excitation
of channel two from the initial state $|g \rangle$ is not possible
and that
the effectively excited energy interval
$(\overline{\epsilon} - b, \overline{\epsilon} + b)$
also covers continuum states of channel one. 
The early stages of this ionization process will be governed by
an exponential decay of the initial state
$|g\rangle$ with the laser-induced depletion rate $\gamma$,
by autoionization of the excited Rydberg states of channel one
into channel two, and by direct laser-induced ionization into the 
continuum states of channel one.
As long as stochastic ionization is negligible, i.e. for
interaction times $t$ with
$1/\gamma < t < t_c$, this
ionization process will reach a metastable regime. 
Thereby the probability
of ionizing into channel one 
is simply determined by the
part of the effectively excited energy interval 
$(\overline{\epsilon} - b, \overline{\epsilon} + b)$
which is located above the ionization threshold,
$\epsilon_1$,
of channel one.
However, as soon as 
$t > t_c$ it is expected that
the branching ratio between channels one and two
is changed. For interaction times $t>t_c$
all Rydberg states whose autoionization lifetimes exceed the stochastic
ionization time, i.e.
$1/\Gamma_n > t_c$ 
($\Gamma_n$ is the autoionization rate of Rydberg state $|n,1\rangle$),
will no longer autoionize into channel two but will
eventually ionize stochastically into channel one. Thus for interaction
times $t > t_c$ it is expected that the probability of ionizing
into channel one is determined by the part of the effectively excited
energy interval
$(\overline{\epsilon} - b, \overline{\epsilon} + b)$
which is located above an energy of the order of $\epsilon_1 - 1/t_c$.
Thus stochastic ionization is expected to
lead to an effective lowering
of the ionization threshold $\epsilon_1$ of channel one.
This manifestation of the competition between autoionization and
stochastic ionization is clearly apparent from Fig. \ref{auto}
where the time evolution of $P_{gg}(t)$ is depicted together with
the corresponding time evolutions of 
$P_{ion-ch1}(t)$ and $P_{ion-ch2}(t)$.
In the case depicted in Fig. \ref{auto}
the laser induced depletion rate $\gamma$ is so small that no
electronic Rydberg wave packet is prepared by power broadening.
However, due to the large laser bandwidth, i.e. $bT \gg 1$,
many Rydberg states are involved in the excitation process.
This implies that to a good degree of approximation
initially state $| g \rangle$ decays exponentially
with rate $\gamma$.

Financial support is acknowledged by
the Deutsche Forschungsgemeinschaft within the Schwerpunktprogramm
`Zeitabh\"angige Ph\"anomene  in Quantensystemen der Physik und Chemie'

\subsubsection*{References}
Agarwal, G. S. (1976) Phys. Rev. Lett. {\bf 37}, 1383.\\
Alber, G., and Zoller, P. (1988) Phys. Rev. A {\bf 37}, 377. \\
Alber, G. (1989) Z. Phys. D {\bf 14}, 307.\\
Alber, G., and Zoller, P.  (1991) Phys. Rep. {\bf 199}, 231.\\
Alber, G. (1992) Phys. Rev. Lett. {\bf 69}, 3045.\\
Alber, G., and Strunz, W. T. (1994), Phys. Rev. A {\bf 50}, R3577.\\
Alber, G., Strunz, W. T., and Zobay, O. (1994)
Mod. Phys. Lett. B {\bf 8}, 1461.\\
Alber, G., and Eggers, B. (1997) Phys.Rev. A {\bf 56}, 820.\\
Aymar, M., Greene, C. H., and Luc-Koenig, E.  (1996)
Rev. Mod. Phys. {\bf 68}, 1015.\\
Beims, M. W., and Alber, G. (1993) Phys.Rev. A {\bf 48}, 3123.\\ 
Beims, M. W., and Alber, G. (1996) J. Phys. B {\bf 29}, 4139.\\ 
Braginsky, V. B., and Khalili, F. Ya. (1992)
{\em Quantum Measurement}, Cambridge University Press, Cambridge.\\
Cooke, W. E., Gallagher, T. F.,
Edelstein, S. A., and Hill, R. M. (1978) 
Phys. Rev. Lett. {\bf 40}, 178. \\
Dando, P. A., Monteiro, T. S., Delande, D., and Taylor, K. T.  (1995)
Phys.Rev. Lett {\bf 74}, 1099. \\ 
Delos, J. B.  (1986) Adv. Chem. Phys. {\bf 65}, 161.\\
Dixit, S. N., Zoller, P., and Lambropoulos, P. (1980)
Phys. Rev. A. {\bf 21}, 1289.\\
van Druten, N. J., and Muller, H. G. (1995)
Phys. Rev. A {\bf 52}, 3047.\\
Eggers, B., and Alber, G. (1998) (in preparation)\\
Fano, U.,  and Rau, A. R. P. (1986)
{\em Atomic Collision and Spectra},
Academic, New York.\\
Gallagher, T. (1994) {\it Rydberg Atoms},
Cambridge University Press, Cambridge.\\
Garraway, B. M., and Suominen, K. A. (1995)
Rep. Prog. Phys. {\bf 58}, 365.\\
Giusti-Suzor, A., and Zoller, P.  (1987) Phys. Rev. A {\bf 36}, 5178.\\
Haken, H.(1970) in {\em Handuch der Physik} (S. Fl\"ugge, ed.)
Vol. XXV/2c, Springer, Berlin.\\
Grobe, R., and Eberly, J. H. (1993) Phys. Rev. A {\bf 48}, 623 (1993)\\
Hanson, L. G., and Lambropoulos, P. (1995) Phys. Rev. Lett. {\bf 74},
5009.\\
H\"upper, B.,  Main, J., and Wunner, G.  (1995)
Phys. Rev. Lett. {\bf 74}, 2650.\\
Jones, R. R., and Bucksbaum, P. H. (1991) Phys. Rev. Lett. {\bf 67},
3215\\ 
Kl\"oden, P. E., and Platen, E. (1992) {\em Numerical Solution of
Stochastic Differential Equations}, Springer, Berlin.\\
Knospe, O., and Schmidt, R.  (1996) Phys. Rev. A. {\bf 54}, 1154.\\
Koch, M., von Plessen, G., Feldmann, J., and Goebel, E. O. (1996)
J. Chem. Phys. {\bf 120}, 367.\\
Landau, L. D., and Lifshitz, E. M. (1975) {\em The Classical Theory
of Fields}, p. 181ff, Pergamon, Oxford.\\
Lankhuijzen, G. M., and Noordam, L. D. (1996) Phys. Rev. Lett. {\bf 76},
1784.\\
Main, J., Wiebusch, G., Welge, K. H., Shaw, J., and Delos, J. B. (1994)
Phys. Rev. A {\bf 49}, 847.\\
Moser, I., Mota- Furtado, F., O'Mahony, P. F., and
dos Santos, J. P.  (1997) Phys. Rev. A {\bf 55}, 3724.\\
Maslov, V. P., and Fedoriuk, M. V. (1981)
{\em Semiclassical Approximation
in Quantum Mechanics}, Reidel, Boston.\\
Mollow, B. R. (1975) Phys. Rev. A {\bf12}, 1919.\\
Robicheaux, F. (1993) Phys. Rev. A {\bf 47}, 1391.\\
Seaton, M. J.(1983) Rep. Prog. Phys. {\bf 46}, 167.\\
Sepulveda, M. A., and Grossmann, F. (1996)
Adv. Chem. Phys. {\bf XCVI}, 191.\\
Stapelfeldt, H., Papaioannou, D. G., Noordam, L. D., and
Gallagher, T. F. (1991) Phys. Rev. Lett. {\bf 67}, 3223\\ 
Vemuri, G., Anderson, M. H., Cooper, J.,
and Smith, S. J. (1991) Phys. Rev. A {\bf 44}, 7635.\\
Zobay, O., and Alber, G.  (1995) Phys. Rev. A {\bf 52}, 541.\\
Zobay, O., and Alber,G. (1996) Phys. Rev. A {\bf 54}, 5361. \\
Zobay, O., and Alber, G. (1998) Prog. Phys. {\bf 46}, 3.\\

\newpage
\begin{figure}[semi]
\caption{\label{semi} 
Schematic representation of the
characteristic spatial regions 
which determine the dynamics of a Rydberg electron. Some classical
trajectories which are relevant for
the semiclassical  wave function are
also indicated.}
\end{figure}

\begin{figure}[ICE]
\caption{\label{ICE} Schematic representation of
a laser-induced isolated core excitation process
in Mg. After initial preparation in a $|3snd\rangle$ Rydberg state
a second laser pulse excites the core $3s \to 3p$ transition.
The Rydberg states of the excited core autoionize.
}
\end{figure}

\begin{figure}[8]
\caption{\label{8} 
Three-channel excitation scheme including
spontaneous emission process and autoionization.}
\end{figure}

\begin{figure}[10]
\caption{\label{10} 
Autoionization and resonant excitation of the core
under the condition of period matching , i.e. $T_{orb}=T_{Rabi}$.
The parameters are
$\kappa^{-1}=7{\rm ns}$,
$\overline{\nu}_1 = [-2(\epsilon - \epsilon_{c1})]^{-1/2}= 73$
($T_{orb}=59{\rm ps}$),  
$\mu_1 = 0.0$, $\mu_2 = 0.5 + i0.1$, $\tau_a = 0.4T_{orb}$
with ${\cal E}_a(t) =
{\cal E}^{(0)}_a e^{-4(t - t_a)^2{\rm ln}/\tau_a^2}$.
Fig. 4a: Scaled ionization rate
$\tilde{\gamma}(t) = \gamma(t)
T_{orb} \tau_a/\mid {\cal D}_{g {\bf e}_a}^{(-)}
{\cal E}_a^{(0)} \mid^2$ as obtained from the optical Bloch equations
(full curve);
Figs. 4b and 4c: Scaled zero- and one-photon contributions
$\tilde{\gamma}^{(0)}(t)$ and  
$\tilde{\gamma}^{(1)}(t)$.
(Reprinted from Zobay and Alber (1996), copyright 1998 by the
Amercian Physical Society)}
\end{figure}

\begin{figure}[two]
\caption{\label{two} Excitation of an isolated Rydberg state:
Radial contour plot (a) and $P_{gg}(t)$,
$P_{ion}(t)$ (b)
as a function of the interaction time $t$ in units of the mean
Kepler period $T$.
The parameters are
$\overline{n}= (-2\overline{\epsilon})^{-1/2}=80$ ($T = 78{\rm ps}$),
$\gamma T = 0.1$, $b T = 0.01$.
Various
approximate asymptotic time dependences are also indicated,
namely Eq.(\ref{1}) (short dashed) and
Eqs.(\ref{2}) and  (\ref{ion}) (long dashed).
}
\end{figure}

\begin{figure}[wave]
\caption{\label{wave} Excitation of an electronic Rydberg wave packet by
laser-induced power broadening:
Radial contour plot (a) and $P_{gg}(t)$,
$P_{ion}(t)$ (b)
as a function of the interaction time $t$ in units of the mean
Kepler period $T$.
The parameters are
$\overline{n}= (-2\overline{\epsilon})^{-1/2}=80$ ($T = 78{\rm ps}$),
$\gamma T = 10.0$, $b T = 10.0$.
Various
approximate asymptotic time dependences are also indicated,
namely Eq.(\ref{1}) (short dashed) and
Eqs.(\ref{2}) and  (\ref{ion}) (long dashed).
$\gamma t = 10.0$, $b T = 10.0$.
}
\end{figure}

\begin{figure}[auto]
\caption{\label{auto} Competition between autoionization
and stochastic ionization:
Time evolution of $P_{gg}(t)$ and of the ionization
probabilites into channels one and two 
$P_{ion-ch1}(t)$ and $P_{ion-ch2}(t)$.
The parameters are
$\overline{n} = \alpha_1 + (-2\overline{\epsilon})^{-1/2} = 80$,
$\alpha_1 = 0.1$,
$\gamma T = 1.0$, $bT = 300.0$,
$\Gamma_n = 2\tau(n - \alpha_1)^{-3}/\pi$
with $\tau = 10^{-5}$ a.u.~.}
\end{figure}

\begin{figure}
\centerline{\psfig{file=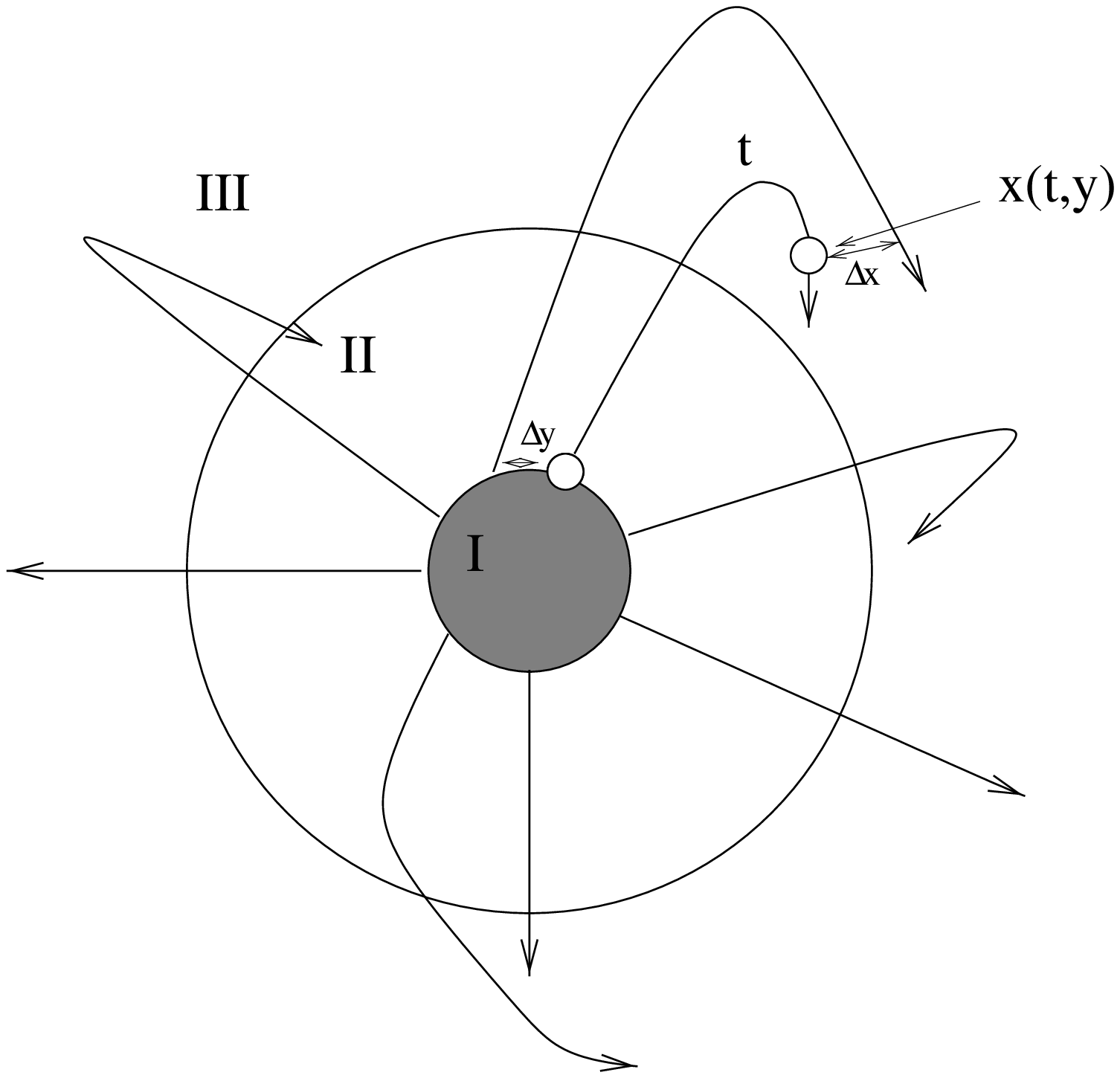,width=4.0in}}
\end{figure}
\begin{figure}
\centerline{\psfig{file=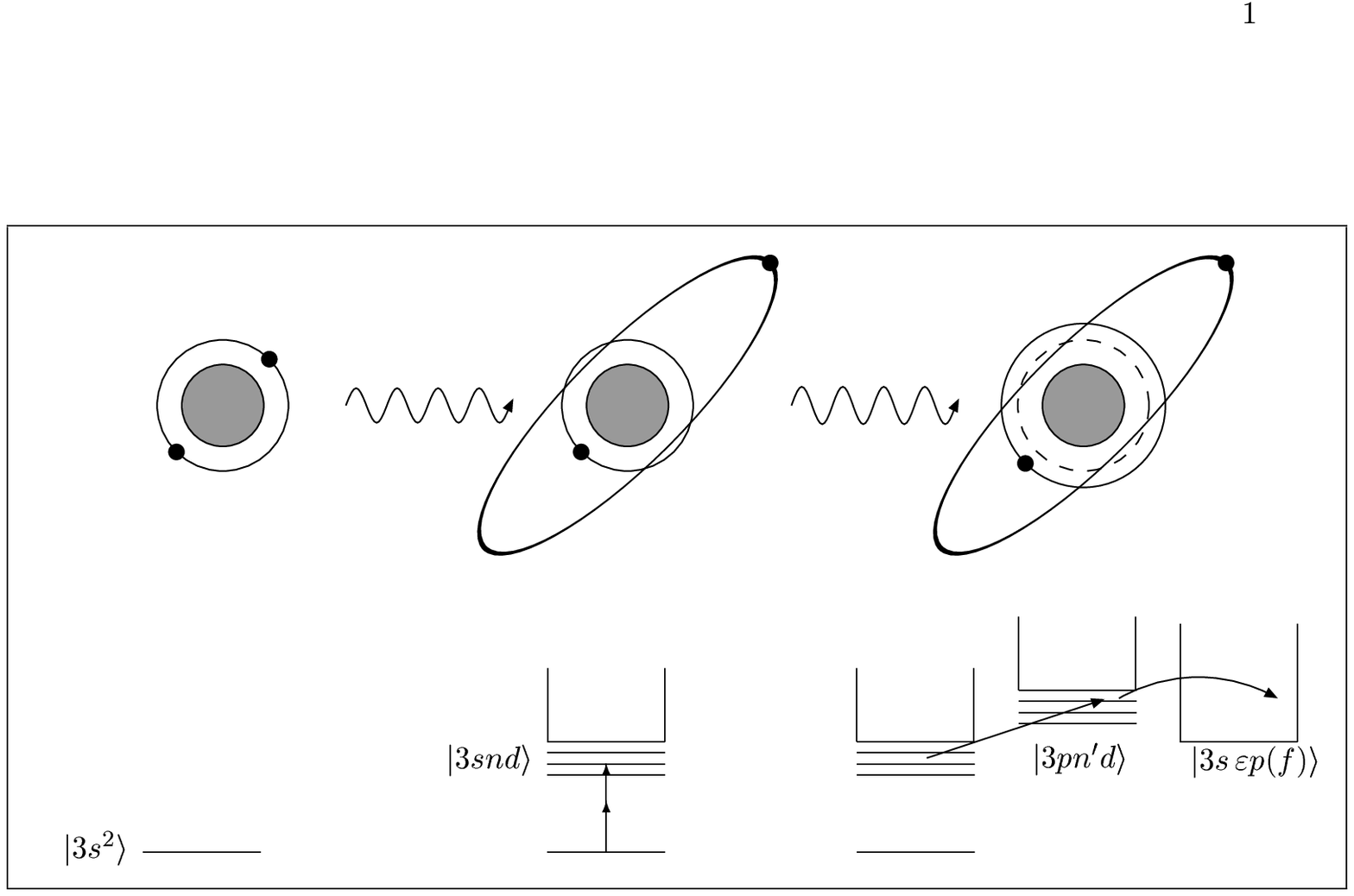,width=4.0in}}
\end{figure}
\begin{figure}
\centerline{\psfig{file=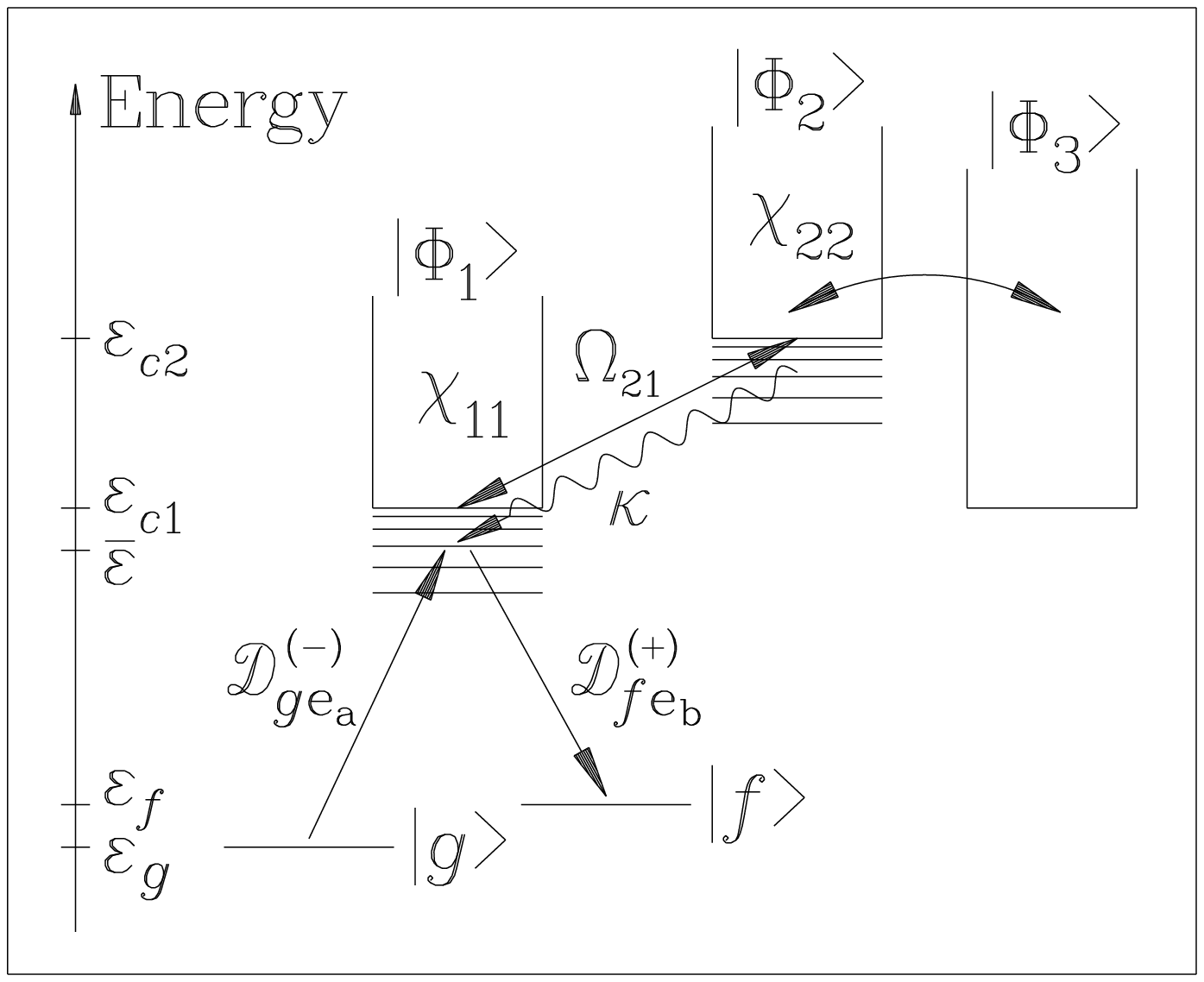,width=4.0in}}
\end{figure}
\begin{figure}
\centerline{\psfig{file=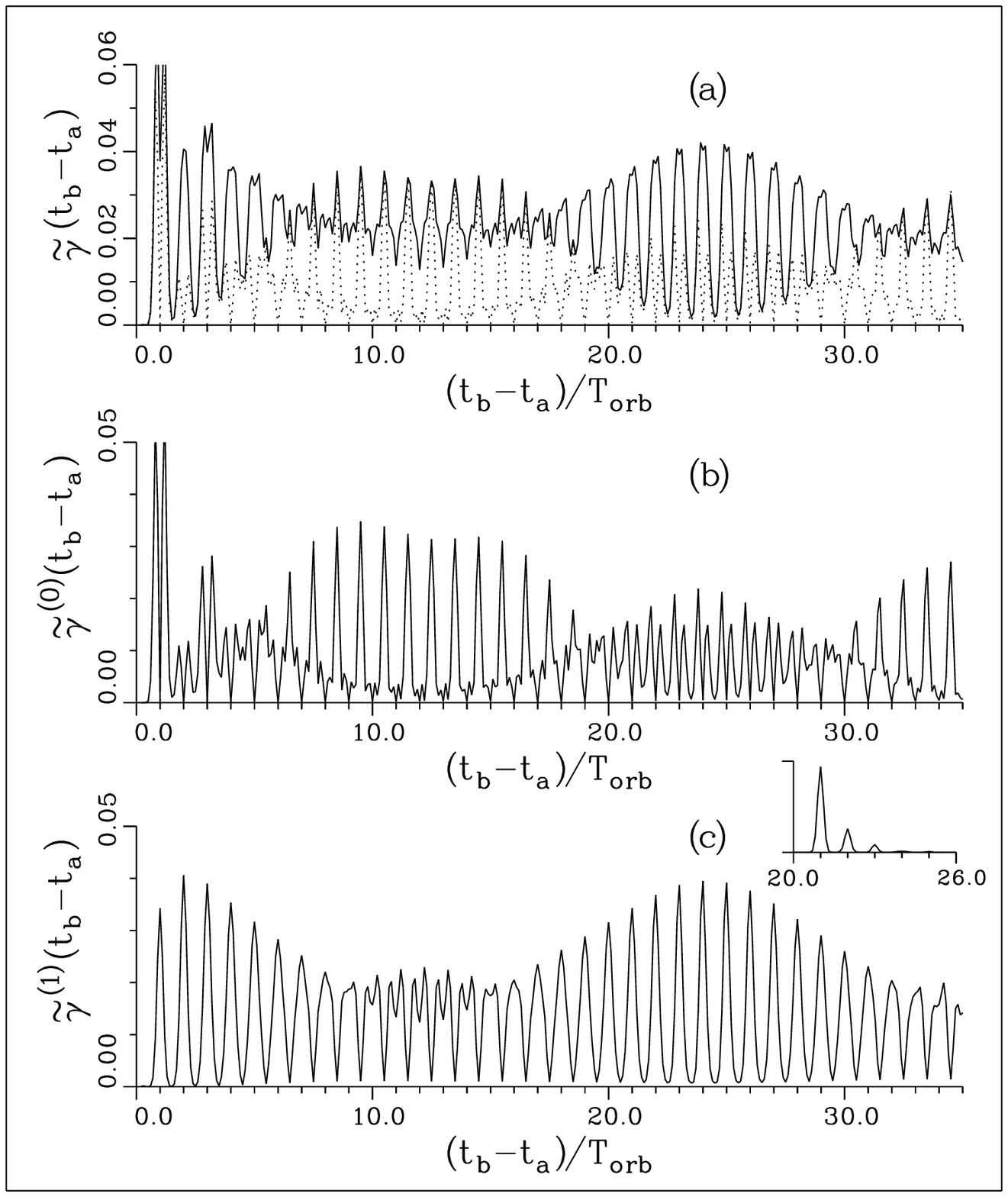,width=4.0in}}
\end{figure}
\begin{figure}
\centerline{\psfig{file=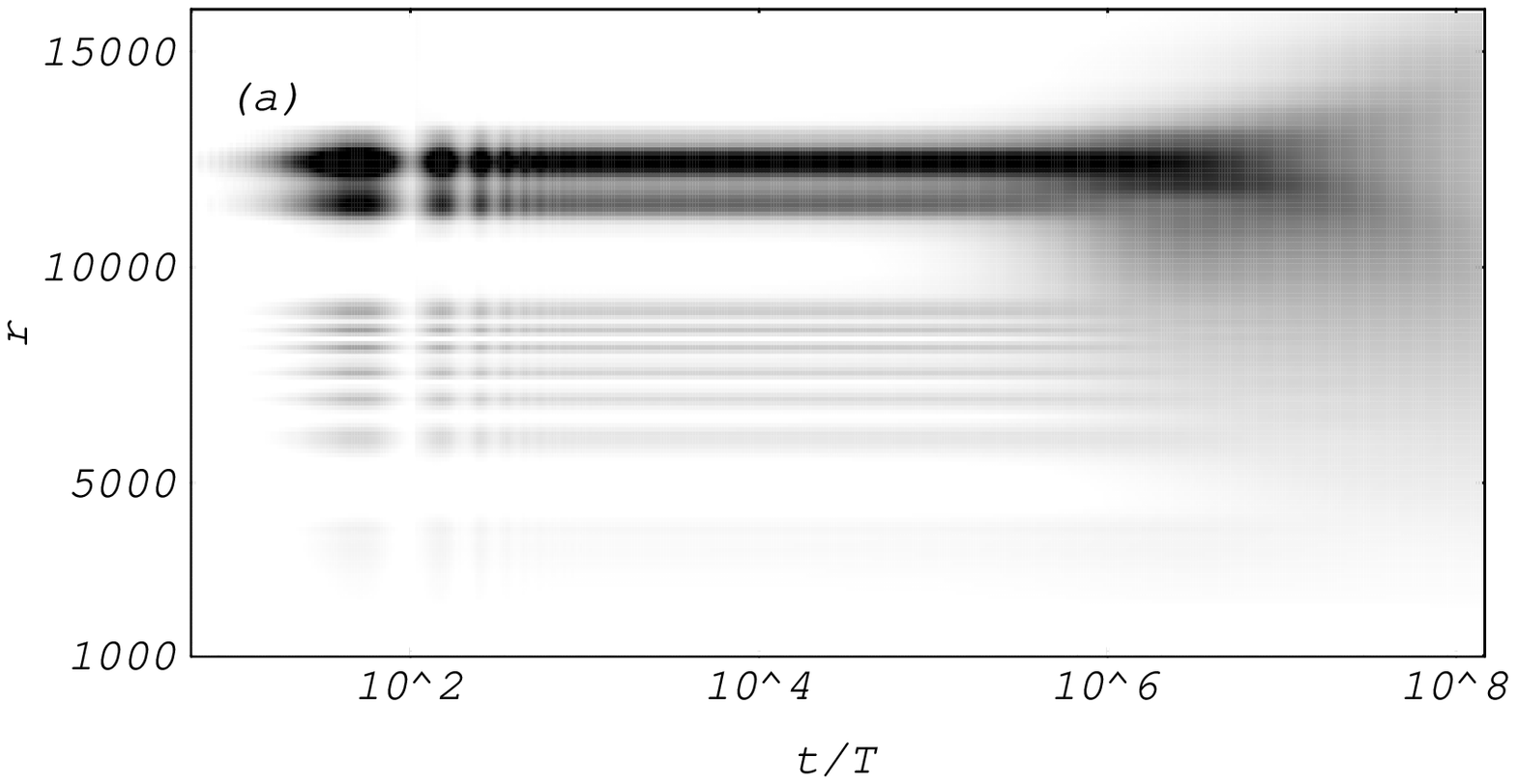,width=4.0in}}
\end{figure}
\begin{figure}
\centerline{\psfig{file=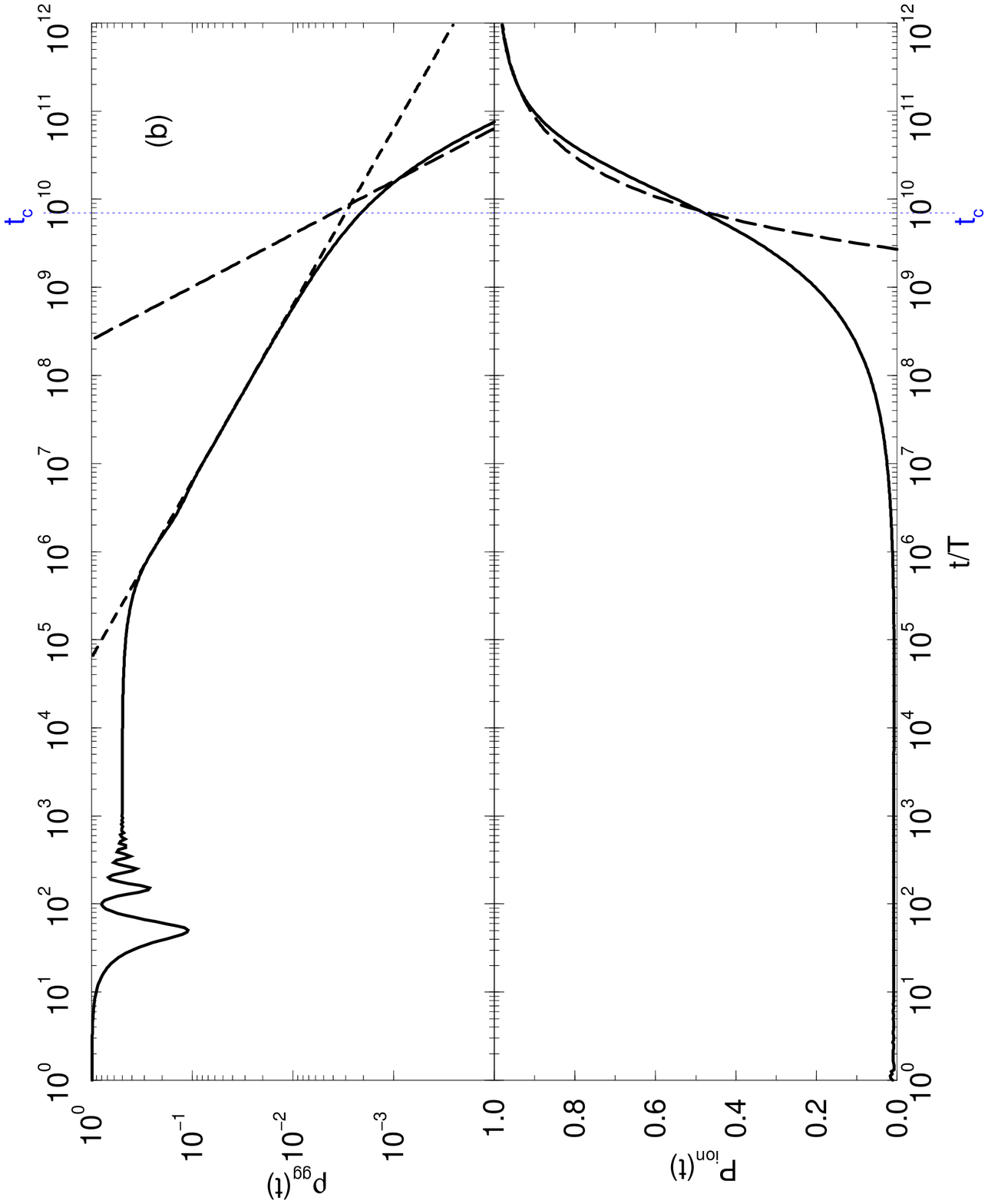,width=4.0in}}
\end{figure}
\begin{figure}
\centerline{\psfig{file=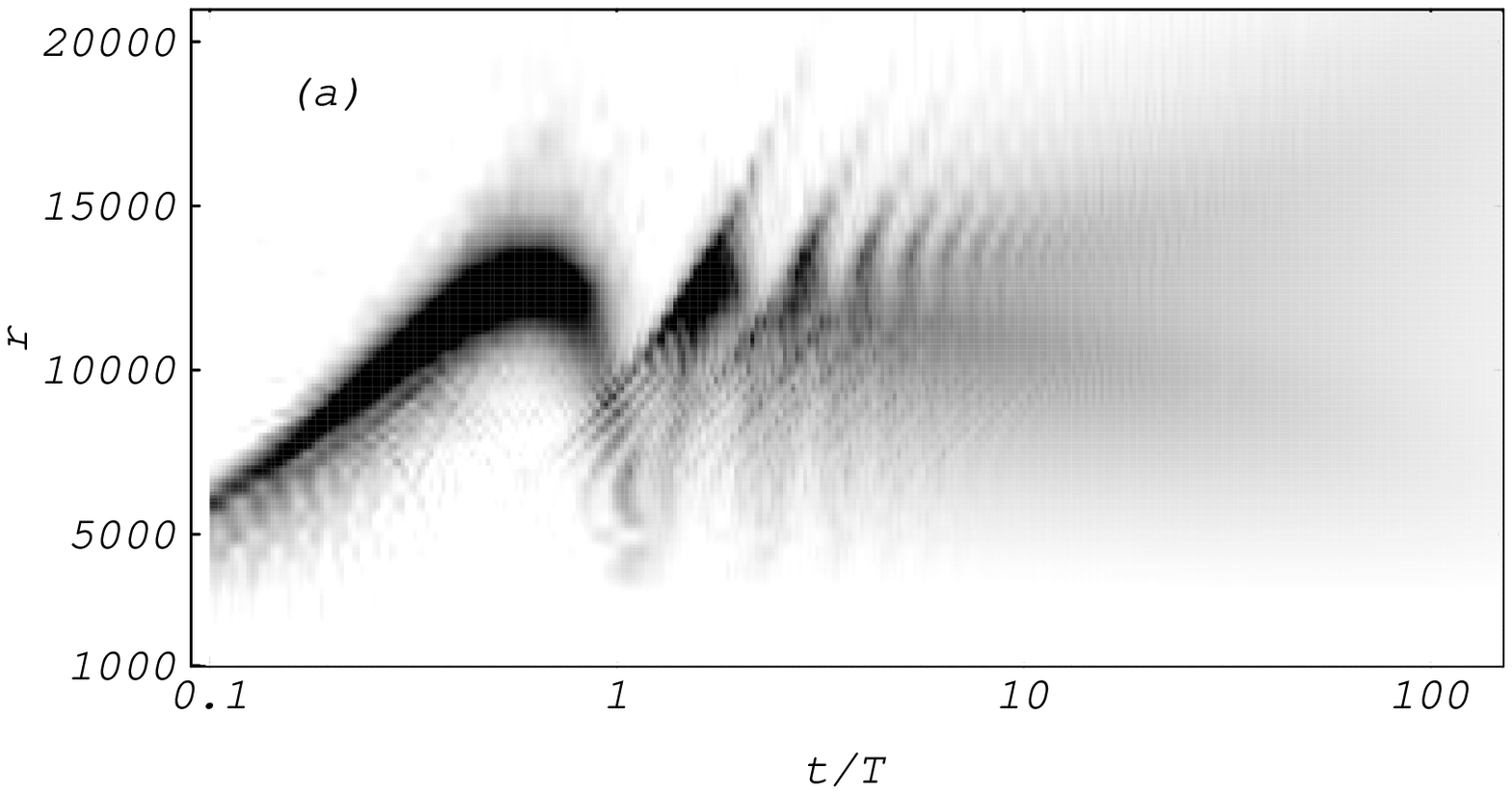,width=6.0in}}
\end{figure}
\begin{figure}
\centerline{\psfig{file=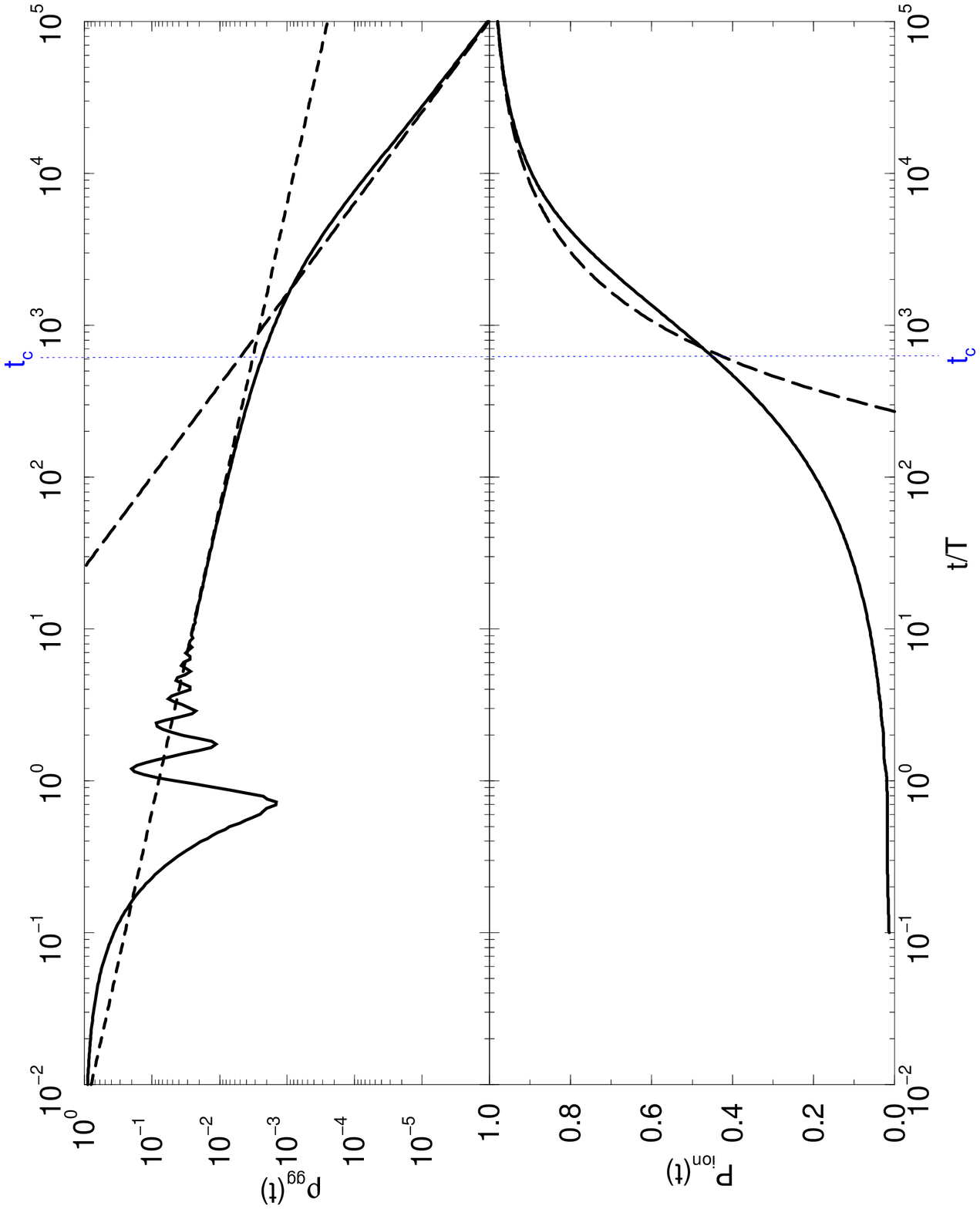,width=6.0in}}
\end{figure}
\begin{figure}
\centerline{\psfig{file=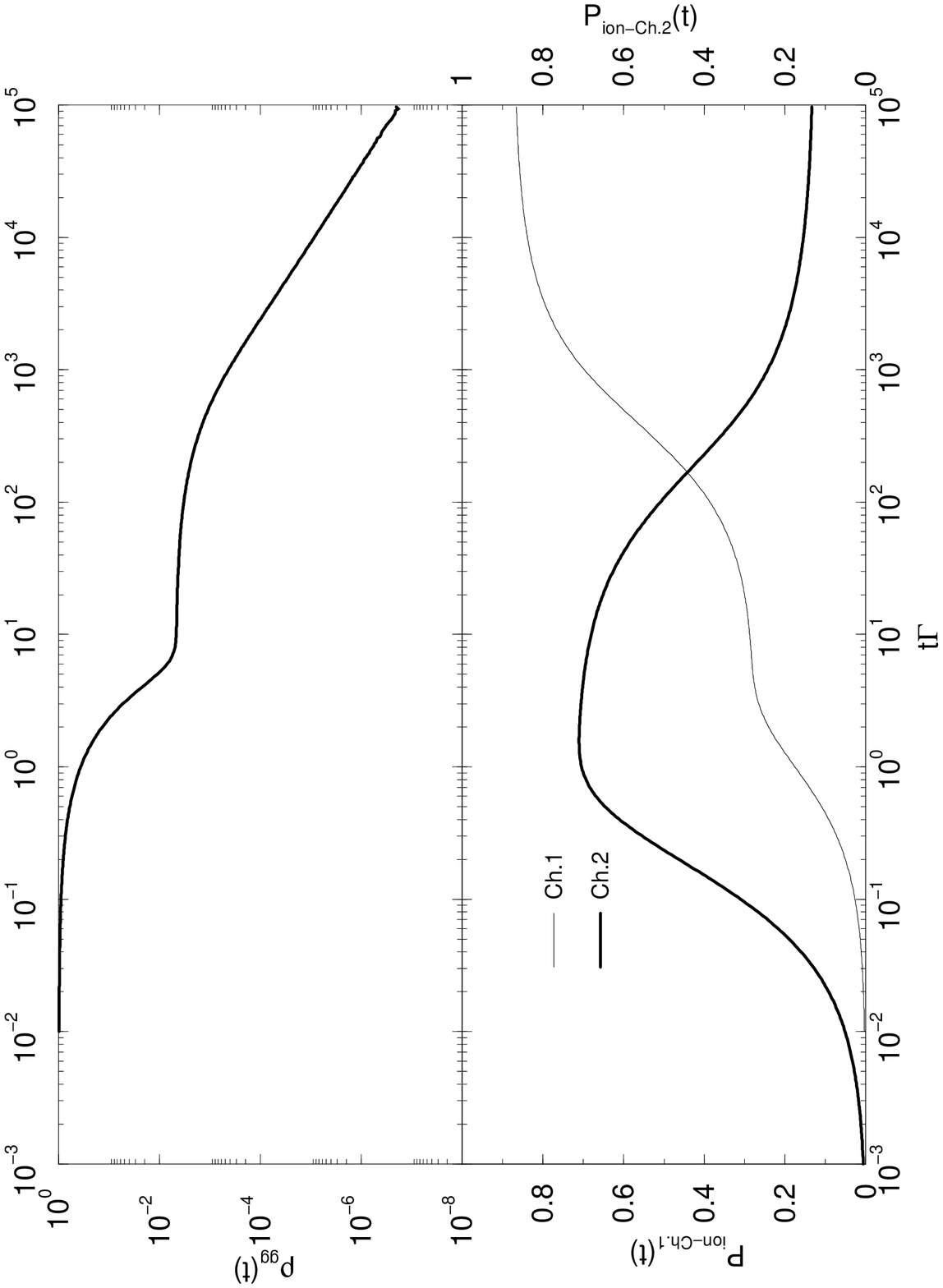,width=6.0in}}
\end{figure}
\end{document}